\documentclass[sn-mathphys-num]{sn-jnl}

\usepackage{array} 
\usepackage{graphicx}%
\usepackage{multirow}%
\usepackage{amsmath,amssymb,amsfonts}%
\usepackage{amsthm}%
\usepackage{mathrsfs}%
\usepackage[title]{appendix}%
\usepackage{xcolor}%
\usepackage{textcomp}%
\usepackage{manyfoot}%
\usepackage{booktabs}%
\usepackage{algorithm}%
\usepackage{algorithmicx}%
\usepackage{algpseudocode}%
\usepackage{listings}%

\usepackage[utf8]{inputenc} 
\usepackage[T1]{fontenc}    
\usepackage{url}            
\usepackage{booktabs}       
\usepackage{amsfonts}       
\usepackage{amsmath}
\usepackage{nicefrac}       
\usepackage{microtype}      
\usepackage{xcolor}         
\usepackage{multirow}
\usepackage{graphicx}
\usepackage{subcaption} 
\usepackage[export]{adjustbox}
\usepackage{enumitem}
\usepackage{breakcites}
\usepackage{tikz}
\usepackage{comment}

\theoremstyle{thmstyleone}%
\theoremstyle{thmstyletwo}%

\theoremstyle{thmstylethree}%

\begin{document}

\title[Integrating Deep Learning in Cardiology: A Comprehensive Review of Atrial Fibrillation Detection, Left Atrial Scar Segmentation, and the Frontiers of State-of-the-Art Techniques]{Integrating Deep Learning in Cardiology: A Comprehensive Review of Atrial Fibrillation, Left Atrial Scar Segmentation, and the Frontiers of State-of-the-Art Techniques}






\author{\fnm{Malitha} \sur{Gunawardhana}}
\author{\fnm{Anuradha} \sur{Kulathilaka}}
\author{\fnm{Jichao} \sur{Zhao}}

\affil{\orgdiv{Auckland Bioengineering Institute}, \orgname{University of Auckland}, \orgaddress{\country{New Zealand}}}


\abstract{Atrial fibrillation (AFib) is the prominent cardiac arrhythmia in the world. It affects mostly the elderly population, with potential consequences such as stroke and heart failure in the absence of necessary treatments as soon as possible. The importance of atrial scarring in the development and progression of AFib has gained recognition, positioning late gadolinium-enhanced magnetic resonance imaging (LGE-MRI) as a crucial technique for the non-invasive evaluation of atrial scar tissue. This review delves into the recent progress in segmenting atrial scars using LGE-MRIs, emphasizing the importance of precise scar measurement in the treatment and management of AFib. Initially, it provides a detailed examination of AFib. Subsequently, it explores the application of deep learning in this domain. The review culminates in a discussion of the latest research advancements in atrial scar segmentation using deep learning methods. By offering a thorough analysis of current technologies and their impact on AFib management strategies, this review highlights the integral role of deep learning in enhancing atrial scar segmentation and its implications for future therapeutic approaches.}




\maketitle

\section{Introduction}
\begin{figure}
    \centering
    \includegraphics[width=0.75\linewidth]{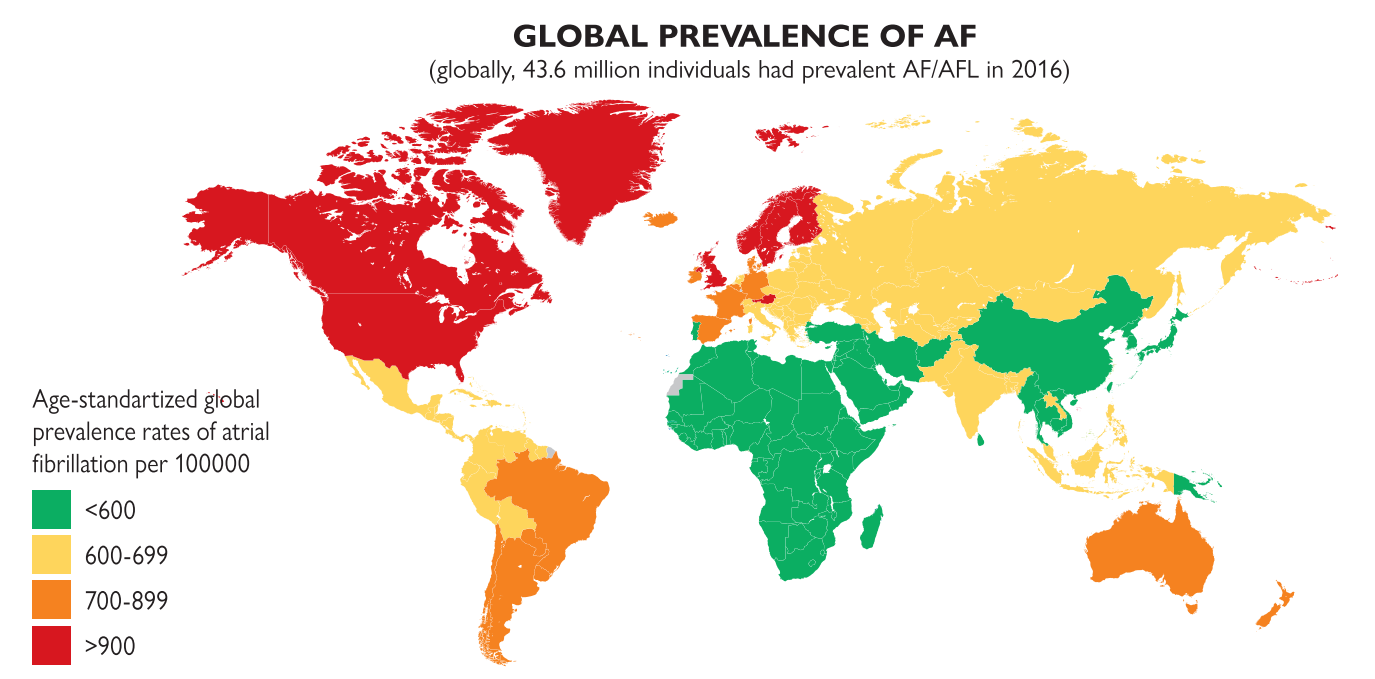}
    \caption{Global prevalence of atrial fibrillation. The image is adapted from \cite{hindricks20212020}.}
    \label{fig:global_map}
\end{figure}

\begin{figure}
    \centering
    \includegraphics[width=1\linewidth]{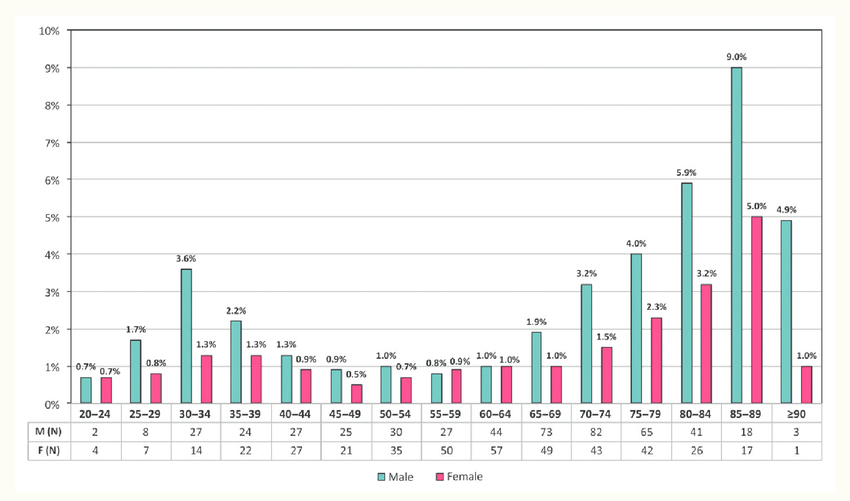}
    \caption{Prevalence of atrial fibrillation by Gender and Age in 65,747 subjects. Individuals screened in Belgium during the week of the heart rhythm, 2010-2014. Data includes males (M) and females (F), presented with a 95\% confidence interval. Image is taken from \cite{mairesse2017screening}}
    \label{fig:male_vs_female}
\end{figure}

Atrial fibrillation (AFib) is the most common cardiac arrhythmia in the world. It is a rapid and irregular heart rhythm, primarily affecting the atria, and  Benjamin et al.\cite{benjamin2019heart} define it as "\emph{supraventricular tachyarrhythmia with uncoordinated atrial electrical activation and consequently ineffective atrial contraction}". This arrhythmia can result in the formation of blood clots within the heart, posing severe health risks, including stroke and heart failure. The implications of AFib are particularly concerning due to its association with increased morbidity and mortality rates, especially among elderly individuals. Given the global demographic shift towards an ageing population, especially among those over 80 years old, AFib is anticipated to become a more prevalent and pressing healthcare issue. The lifetime risk of AFib is approximately one in three to five individuals beyond the age of 45 \cite{linz2024atrial}, and the prevalence of AFib can reach up to 18\% in individuals aged 80 years and older \cite{teh2023atrial}. The global prevalence of AFib has risen from 33.5 million to 59 million individuals between 2010 and 2019. The prevalence of AFib is expected to rise to 15.9 million in 2050 in America and 17.9 million in 2060 in Europe \cite{linz2024atrial}. According to Fig.~\ref{fig:global_map}, North America is suffering more from AFib compared to other countries, and in the United States, it is estimated that 12.1 million people will have AFib in 2030 \cite{okoli2024regional}. Also, compared to women, AFib is more common in males (Fig.~\ref{fig:male_vs_female}) \cite{benjamin2019heart} and comparing the number of patients in the USA and other countries, the United States had the highest age-standardized rate of incidence (ASIR) and increased more quickly than other countries. China carried the largest burden of AFib, followed by the United States and India \cite{ma2024global}.

AFib’s clinical presentation can vary. While some individuals may experience no symptoms, others might suffer from palpitations, dyspnea, or lightheadedness. The nature of AFib episodes can also differ, with some being transient and others more persistent. Despite not being inherently life-threatening, AFib demands prompt and appropriate medical intervention to avert complications.

Fibrosis is a critical pathological change in the heart that plays a major role as an arrhythmogenic substrate in AFib and it is essentially a form of scarring that occurs within the heart tissue. When the heart tissue is damaged, the body initiates a healing response, which involves the activation of fibroblasts. These cells are responsible for producing the extracellular matrix components that form the scar tissue~\cite{chen2013fibroblasts}. Over time, excessive activation of fibroblasts can lead to an abnormal buildup of fibrous tissue, replacing the normal myocardial cells. This excessive fibrous tissue known scar tissue, which is less elastic and more rigid compared to the normal cardiac tissue~\cite{humeres2019fibroblasts}. The presence of fibrotic scar tissue can lead to heterogeneity in the conduction properties of the atrial tissue. Parts of the atria with dense fibrosis may conduct electrical signals slower than less affected areas. This differential in conduction speeds can facilitate the formation and maintenance of reentrant circuits, which are critical in sustaining AFib. The structural changes due to fibrosis and scarring can also physically alter the geometry of the atria. Such alterations can exacerbate the electrical disturbances by affecting the normal atrial architecture, which is essential for orderly impulse conduction.

Magnetic Resonance Imaging (MRI) technology enhancement, mainly through contrast agents like Gadolinium, has led to the development of late gadolinium-enhanced MRI (LGE-MRI). This advancement has been instrumental in detecting scar tissue. Studies at the University of Utah have shown a correlation between fibrosis in the left atrial (LA) wall and outcomes in AFib ablation procedures  \cite{oakes2009detection,akoum2013association}. LGE-MRI has become the preferred technique for detecting and quantifying atrial wall scar tissue. Accurate atrial segmentation and scar quantification are essential for the effective treatment of AFib. However, the LA cavity, characterised by its small volume and complex anatomy, is challenging to segment due to the thin atrial wall and the anatomical structures around the atria that exhibit similar intensities, which can confuse segmentation algorithms \cite{wang2019robust,maceira2010reference}. Consequently, manual segmentation remains a problematic and error-prone process \cite{tobon2015benchmark}. In contrast, the advent of deep learning offers a promising alternative. The capability of pattern recognition and predictive analytics could enable more accurate prognoses of AFib, potentially improving patient outcomes. This technological advancement can alleviate healthcare professionals’ therapeutic workload, providing a more streamlined, scalable, and precise method for managing this increasingly prevalent cardiac condition.

The rising interest in deep learning within research challenges underscores a paradigm shift in atrial segmentation and clinical imaging development. The trend towards deep learning-based approaches is poised to revolutionise clinical practices in the coming years. The growing importance of atrial segmentation and quantifying scar tissues is evident, and deep learning plays an increasingly significant role. For instance, in the LAScarQS 2022: Left Atrial and Scar Quantification \& Segmentation Challenge \cite{zhuang2023left}, most methods employed deep learning for segmenting the LA cavity and scars from LGE-MRI images, demonstrating superior performance over traditional methods \cite{pop2019statistical}.

Several comprehensive reviews have been published, shedding light on various aspects of this domain. Notably, Jammart et al.~\cite{jamart2020mini} provide an in-depth exploration of the application of deep learning techniques for atrial segmentation from LGE-MRIs. Wu et al.~\cite{wu2021recent} offer a detailed analysis of fibrosis from cardiac MRI, underscoring the critical role of precise scar identification in the assessment of cardiac health and disease progression. Then Li et al.~\cite{li2022medical} extend this discussion by focusing on medical image analysis tailored to LA LGE-MRI for AFib. Despite these valuable contributions, according to our knowledge, we are the first ones to focus solely on LA scar segmentation using deep learning methods due to its challenging nature. In this paper, we first focus on the mechanism of AFib and the theory and concepts behind MRI and electrocardiogram (ECG/EKG). Then, we shift our attention to the deep learning by providing a comprehensive review of the concepts behind it. Then, we focused our attention on the current state-of-the-art deep learning algorithms in scar segmentation.  Finally, we discussed the future directions for using deep learning in LA scar segmentation. 

Compared to the existing articles as described above, our paper is designed to be accessible to a broad audience, ranging from medical professionals seeking to understand the latest advancements in deep learning in imaging technology to machine learning experts interested in the application of their skills to real-world health challenge in LA scar segmentation. The initial segments of our paper are crafted to provide machine learning experts with a comprehensive understanding of the anatomical and physiological aspects of AFib. In contrast, the latter sections are intended to acquaint medical professionals with the usage of deep learning. This dual focus is a distinctive feature of our work, as we aim to bridge the gap between technical computational methodologies and clinical application. Our comprehensive review and novel insights into the segmentation process are presented in a manner that is both informative for experts in the field and accessible to newcomers. This approach addresses a significant gap in the existing literature, where the intersection of medical knowledge and machine learning expertise is often overlooked.  \emph{Our goal is for this paper to serve as a valuable resource for researchers seeking a holistic understanding of medical and machine learning aspects in the context of LA scar segmentation without needing to consult multiple sources.}

\section{Atrial fibrillation}



In a typical heart, an electrical impulse originates from the sinus node in the right atrium, leading to a heart rate of 60 to 150 beats per minute in a healthy individual. However, in the case of AFib, multiple electrical impulses are generated from various sites within both atria. This results in the atria contracting at an excessively high rate, often exceeding 400 times per minute. Consequently, the ventricles struggle to match this rapid pace of contractions. They tend to beat faster than expected and may not adequately fill with blood or pump it out efficiently. This irregularity causes blood to accumulate in the atria rather than being efficiently transferred to the ventricles and circulated throughout the body. Such blood stagnation heightens the likelihood of blood clots forming within the heart. These clots can then dislodge, enter the bloodstream, and potentially reach the brain, thereby significantly increasing the risk of stroke. Based on the duration of the AFib, there are five types of AFib.

\begin{itemize}
    \item \textit{First diagnosed AFib}:- Identifying AFib on a patient for the first time using ECG despite the duration.  
    \item \textit{Paroxysmal AFib}:- This type of AFib resolves on its own within a week. Usually, this type of AFib ends within two days. 
    \item \textit{Persistent AFib}:- This type of AFib lasts more than a week. In this period, recurrent episodes of AFib can occur.
    \item  \textit{Long-standing persistent AFib}:- This type of AFib is continuous for more than a year. 
    \item \textit{Permanent AFib}:- At this stage, both the patient and physician have acknowledged the AFib after several attempts to return the heartbeat to normal. However, if the patient and clinician agree to restore the sinus rhythm again, this classification could potentially be reconsidered. 
\end{itemize}

\begin{figure}
    \centering
    \includegraphics[width=1\linewidth]{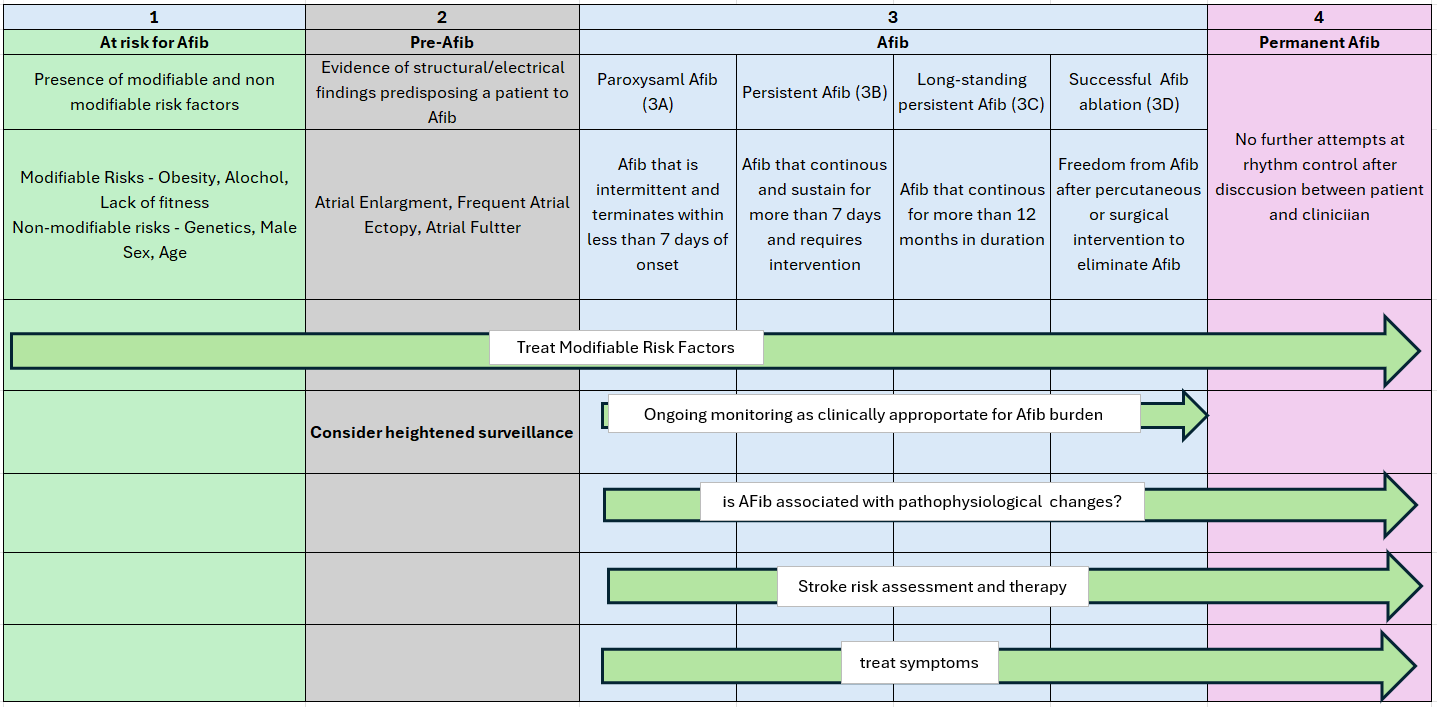}
    \caption{AFib Categorization. In Category 3, patients can transition from different sub-stages. Figure is modified from \cite{joglar20242023} }
    \label{fig:afib_cat}
\end{figure}

Although the above classification based on duration is validated by the guidelines of the European Society of Cardiology (ESC) in 2020 \cite{hindricks20212020}, the guidelines of the Joint Committee of the American College of Cardiology and the American Heart Association (ACC / AHA) in 2023 \cite{joglar20242023} argue that this classification is focused on therapeutic interventions. Therefore, they proposed a new classification, which uses stages and aims to rectify the shortcomings of the previous system by acknowledging AFib as a disease that progresses over time, necessitating varying strategies at its different stages (Fig.~\ref{fig:afib_cat}). This includes everything from prevention and early detection to management of heart rate and rhythm. 

Although AFib causes a serious risk for patients, more than 20\% of the patients have no symptoms to identify. Therefore, AFib is identified when patients are admitted to the hospital due to complications of AFib, such as stroke. The most common symptoms of AFib may include, but are not limited to palpitations (fast, pounding or fluttering heartbeat), dyspnea (difficulty of breathing), fatigue, chest and/or throat discomfort and dizziness.  In addition to heart defects such as congenital heart defects and sick sinus syndrome, heart attacks, hypertension, diabetes, obesity, sleep apnea, alcohol, smoking, genetics, ethnicity, chronic kidney disease, and sometimes excessive sports activities act as risk factors for AFib. \cite{schoonderwoerd2008new, hindricks20212020}

\subsection{Mechanism of Atrial Fibrillation}

Despite extensive research on atrial fibrillation, the precise mechanisms of initiation and continuation of AFib remain unclear. Studies suggest that the initiation and maintenance of AFib involve two critical components: \textit{triggers} and \textit{substrates}. Triggers are specific events or conditions that can precipitate an AFib episode, acting as the initial spark for the arrhythmia. On the other hand, substrates refer to the underlying structural and mechanical characteristics of the atria that allow AFib to occur and continue \cite{olshansky2023mechanisms}. These elements include abnormalities in the atrial tissue that provide a conducive environment for the persistence of the arrhythmia.


As AFib evolves, it typically transitions from being primarily driven by these triggers in its paroxysmal stage to becoming more reliant on the altered substrate for its continuation, leading to persistent AFib. Research has identified two key processes that play a pivotal role in sustaining AFib and contributing to its progression: (1) the presence of one or multiple ectopic atrial foci that fire rapidly, generating irregular fibrillatory activity, and (2) the formation of one or a few re-entrant circuits within the atria. These circuits can cause rapid local depolarization, facilitating the continuation of atrial fibrillatory activity that characterizes AFib.

\subsubsection{Atrial Ectopic Foci}

\begin{wrapfigure}{r}{0.6\textwidth}
  \begin{center}
   \vspace{-5mm}
    \includegraphics[width=0.48\textwidth]{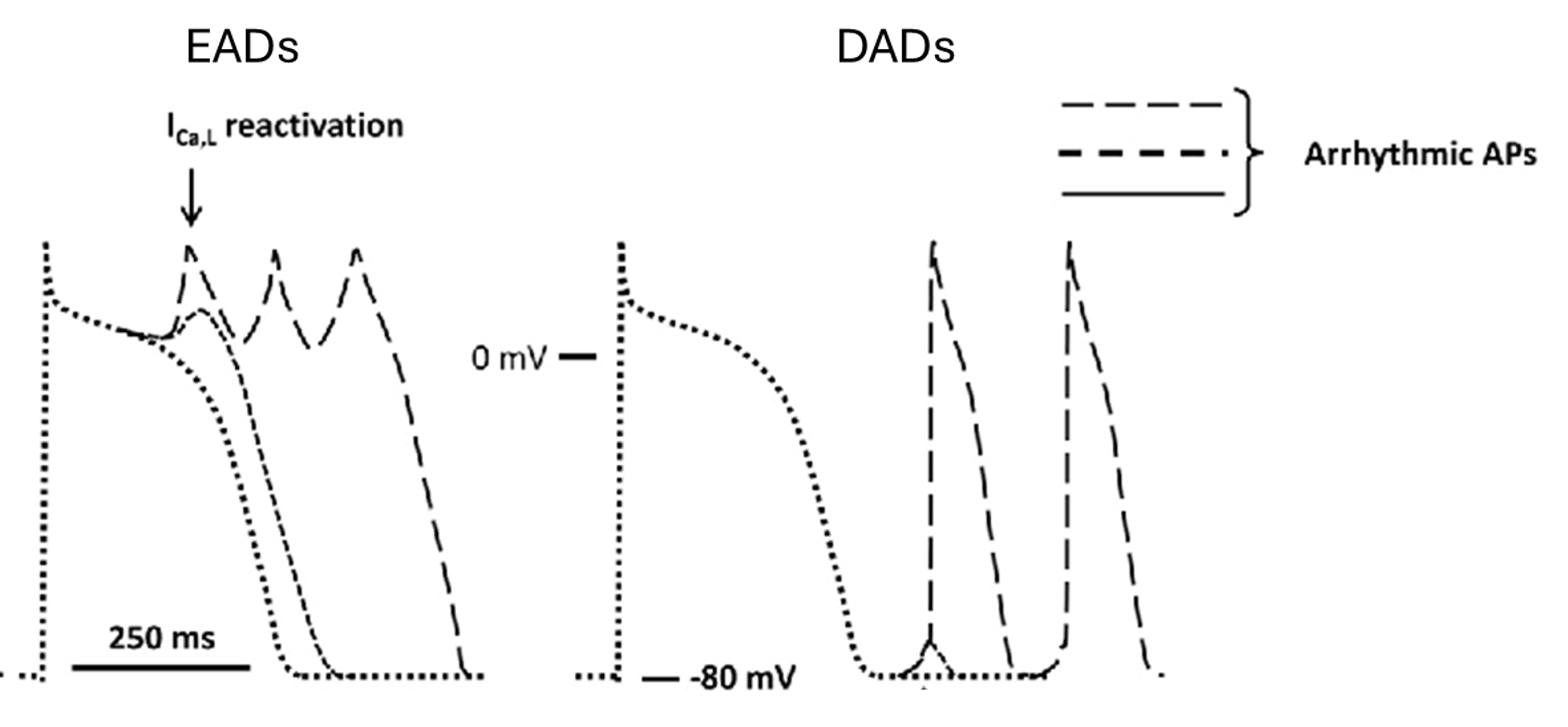}
     \vspace{-2mm}
  \end{center}
  \caption{Representation of EADs and DADs. Image is taken from \cite{iwasaki2011atrial}}
   \vspace{-5mm}
   \label{fig:ead}
\end{wrapfigure}

These are the abnormal firing points within the atria that send out irregular electrical pulses. According to the \cite{calkins2007hrs}, authors find out that 85-95\% of paroxysmal AFib episodes are initiated in pulmonary vein (PV) foci.  There are two main mechanisms that cause ectopic activities. Those are known as early after-depolarization (EAD) and delayed after-depolarisation (DAD). Both of them are shown in Fig.~\ref{fig:ead}.

DADs are abnormal spontaneous depolarizations that occur during phase 4 in cardiomyocytes after the end of normal action potential repolarization. After repolarization, the ryanodine receptor (RyR) on the sarcoplasmic reticulum (SR) membrane is usually closed. However, if the SR is overloaded or the RyR receptor is defective, the latter can release Ca$^{2+}$ from the SR store into the cytoplasm. This Ca$^{2+}$ increase in the cytoplasm leads to the activation of the sodium-calcium exchanger (NCX), exchanging 1Ca$^{2+}$ outward with 3Na$^{+}$ inward, leading to a positive net flux entering the cell, causing an inward current. This inward current is called transient inward current, generating a DAD, and can be sufficient to fire an action potential.

DADs, on the other hand, occur after the heart cell's electrical reset is complete. Normally, $Ca^{2+}$ channels on the storage unit within the cell, the sarcoplasmic reticulum, are shut tight. But, if this storage is too full or the channels (known as ryanodine receptors) are not working right, they leak $Ca^{2+}$ into the cell. This leak prompts a process where the cell swaps out $Ca^{2+}$ for $Na^{+}$ ions, resulting in a net gain of positive charge inside the cell. This shift can spark another heartbeat. DADs are spontaneous electrical pulses that arise during the cell's resting phase, driven by the leakage or release of $Ca^{2+}$ from the sarcoplasmic reticulum, leading to an activation of the $Na^{+}$-$Ca^{2+}$ exchanger mechanism, which creates a net inward current, potentially triggering an additional heartbeat.


\subsubsection{Re-entry}

After triggers ignite the arrhythmia, other factors step in to maintain the AFib. Among them, re-entry is the main factor. It can be defined as a self-perpetuating circuit within the atria. An electrical impulse gets trapped in this circuit, continuously circling and re-exciting the tissue like a runaway train on a never-ending track. 

The re-entry circuit concept, particularly the leading-circle model \cite{allessie1977circus}, illustrates how the heart's electric waves can circle around an area that doesn't respond to electrical activation, known as a non-excitable area. The wave travels around a non-excitable area, which can be a part of the heart damaged by a disease, acting like a scarring or blockage. Because this central zone is stuck in a constant state of recovery, it does not allow any other electrical impulses to pass through, making it a static hurdle. This action is like the electric movement looping in a certain pattern, sending electrical lines in all directions: inside the barrier and to the neighbour's heart muscle, moving in a time rhythm.

This phenomenon can be described by a keyword, 'wavelength,' which is the journey an electrical line makes during the hearth's recharge time. Described by a simple equation WL = CV × RF, where 'WL' is the term for wavelength, 'CV' means how quickly the electrical wave is moving or conduction velocity, and 'RF' describes the refractory period. When the system is lithe and able to cope with the power being sent around, the results stick in the sense that the route being used is active, pointing out that a slight wait for the signal may help sustain this circuit. This course can get packed or get room enough for other signs to drive in, setting the level for some types of stable pulses or, in some heads, ending the line of the circuit, such as the heart muscle not being ready to handle another wave due to its pause-time. The following conditions can fuel this re-entry. 

\begin{itemize}
    \item Slowed conduction: Picture the normal electrical flow as a runner on a clear path. In AFib, the "path" is obstructed, slowing the conduction and making it easier for the re-entry circuit to form.
    \item Shortened refractory period: Imagine the runner needing to rest before sprinting again. The refractory period is the time it takes for heart cells to recover after being activated. In AFib, this "rest" is shortened, allowing impulses to fire prematurely, perpetuating the re-entry loop.
    \item Abnormal calcium handling: Think of calcium as the fuel for the electrical impulses. In AFib, this "fuel" is mishandled, leading to excessive calcium buildup in the cells, further contributing to electrical instability and re-entry.
\end{itemize}

\begin{figure}
    \centering
    \includegraphics[width=0.8\linewidth]{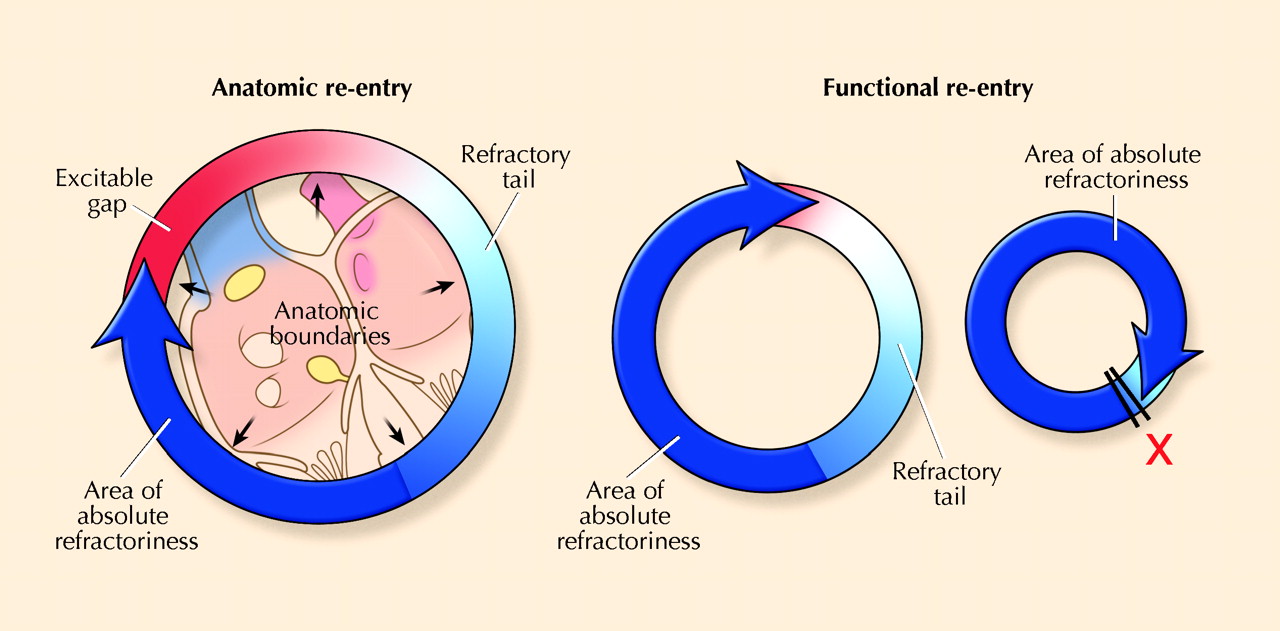}
    \caption{Anatomical re-entry vs functional re-entry. Image is taken from \cite{veenhuyzen2004atrial}}
    \label{fig:reentry}
\end{figure}

Also, there are two main types of re-entries: 1) anatomical re-entry and 2) functional re-entry. Anatomical re-entry is characterized by the development of a re-entrant circuit that navigates around a non-conducting anatomical structure. In contrast, functional re-entry depends on the myocardium's electrophysiological properties to establish the circuit's route. In the case of anatomical re-entry, the circuit's dimensions are dictated by permanent anatomical barriers. Conversely, in functional re-entry, the circuit's dimensions can be calculated as the product of the conduction velocity and the refractory period. If the wavefront progresses too rapidly, or if its refractory period is excessively prolonged, the front part of the wave could collide with its own trailing end, leading to self-extinction. Therefore, these factors critically influence the minimum size that the circuit can maintain (Fig.~\ref{fig:reentry}).

\subsection{Atrial Remodeling}

The development of AFib leads to changes in the heart that further promote AFib, thus creating a cycle of increasing vulnerability to its induction and maintenance. These changes in the atria, which can be electrical, structural, or contractile, are known as atrial remodelling and contribute to conduction disturbances. These disturbances, in turn, increase the likelihood of triggers that are prone to re-entry, such as re-entry-prone substrates. This self-sustaining and escalating nature of AFib is commonly described by the phrase "AFib begets AFib" \cite{wijffels1995atrial}.

\subsubsection{Structural remodeling}

Structural remodelling is a key factor in developing and continuing AFib, involving changes in the heart's physical structure and composition at multiple levels.

\paragraph{At the Cellular Level}

Several processes play a pivotal role in structural remodelling at the cellular level, and \cite{de2011mechanisms} documents two main conditions. Hypertrophy, characterized by the enlargement of heart muscle cells, or cardiomyocytes, occurs in response to increased pressure or volume. This condition is prevalent in both animal studies and patients with AFib, disrupting the conduction of the heart's electrical signals. Concurrently, apoptosis, a regulated form of cell death, has been identified as a significant factor in AFib, suggesting that elevated heart rates might critically influence cell mortality in this context. Additionally, myolysis, the breakdown of muscle fibres, compromises the heart’s contractile capabilities. This phenomenon has been observed in animal models as well as patients suffering from AFib, particularly those experiencing moderate to severe heart failure, as highlighted by \cite{thijssen2001structural}

\paragraph{At the Tissue Level}

AFib primarily induces structural remodelling at the tissue level through the development of atrial fibrosis. This condition arises from an excessive production of extracellular matrix (ECM) collagen within the cardiac connective tissue. Maintenance of cardiac ECM metabolism is facilitated by fibroblasts, which are situated in the interstitial spaces of the myocardium. Cardiac fibrosis represents a maladaptive response in the balance of ECM molecule production, where the synthesis of ECM components significantly surpasses their degradation. This imbalance leads to an unhealthy expansion of connective tissue. The initiation of fibrosis typically serves as a reparative mechanism following tissue injury and cardiomyocyte death, indicating a complex pathophysiological process.

Mechanical, chemical, or electrical disturbances or injuries, with stimuli originating from the ECM, can activate the remodelling process. This process encompasses three distinct phases. The first phase is known as the \textit{inflammatory Phase}. It is triggered by stress signals, and the immune system engages with the cardiac ECM, initiating the production of inflammatory cytokines. This immune response enhances the secretion of matrix metalloproteinases (MMPs) by cardiac fibroblasts, which facilitates the degradation of ECM. The resultant ECM breakdown allows further leukocyte infiltration, marking the beginning of the granulation phase. The second phase known as the \textit{granulation phase} is characterized by the phagocytosis of necrotic cells and ECM remnants, this phase also sees an increased recruitment of fibroblasts through chemotaxis to replenish the degraded ECM, setting the stage for fibrosis development during the scarring phase. In the scarring phase, the recruitment of fibroblasts correlates with a reduction in MMP production and an upsurge in MMP inhibitors, leading to the proliferation and eventual overgrowth of cardiac connective tissue. This overgrowth signifies the transition to a fibrotic state.

As fibrous tissue inherently lacks contractility, its compromised expansibility during systole leads to heightened cardiac wall stiffness and diminished myocardial compliance, adversely affecting cardiac contraction. Additionally, cardiomyocytes may undergo hypertrophy or death, prompting further immune infiltration into the connective tissue and exacerbating fibrosis. Excessive ECM accumulation disrupts intercellular connections, notably the connexins forming gap junctions and impairing cardiac conduction. Concurrently, the fibrotic regions experience diminished oxygen supply, contributing to cellular demise and further cardiac tissue degradation due to fibrogenesis.

\begin{figure}
    \centering
    \includegraphics[width=0.8\linewidth]{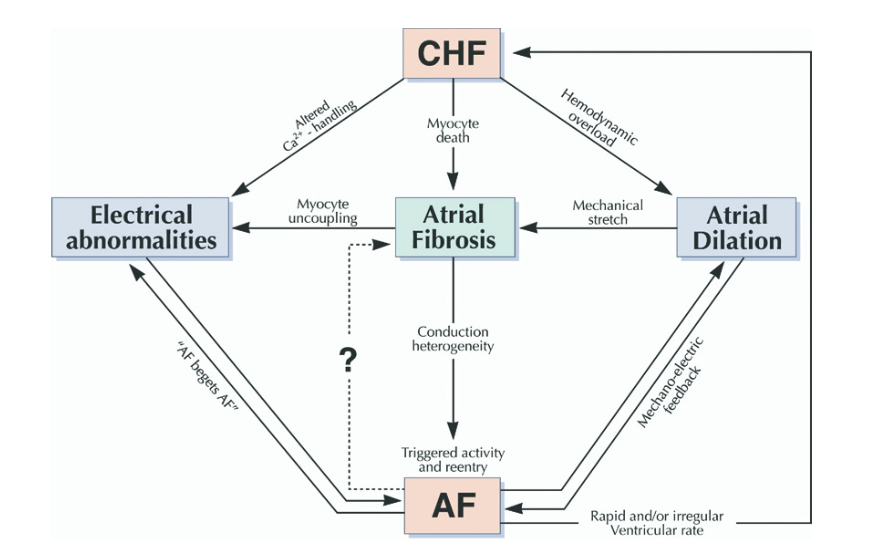}
    \caption{Relationship between AFib, Atrial fibrosis, atrial dilation (enlargement of the left atrium ) and cognitive heart failure (CHF).  AFib often occurs because scar tissue has formed in the heart. This process is called fibrosis. Image is adapted from \cite{burstein2008atrial}}
    \label{fig: afib chf}
\end{figure}

Fibrosis often occurs in patients with AFib and serves as a key indicator of structural changes within the heart. It plays a significant role in the continuation of this arrhythmia by becoming a primary factor for the condition's persistence.  As AFib progresses, becoming more chronic and less amenable to treatment, the spread of fibrosis throughout the heart muscle is noted, which adversely affects the heart's shape, size, and overall function. Consequently, the process of evaluating fibrosis takes on vital importance in the context of diagnosing AFib and determining the potential effectiveness of treatment approaches. In the clinical setting, LGE-MRI (further discussed in Section \ref{mri}) is used to evaluate the extent of fibrosis in the atrial wall, and a study conducted at the University of Utah \cite{mcgann2014atrial} shows a direct connection between AFib treatment outcome and fibrosis. The relationship between fibrosis, AFib, atrial dilation, electrical abnormalities and cognitive heart failures are shown in Fig.~\ref{fig: afib chf}. 

Another outcome of fibrotic changes in the heart's structure is the disruption of normal electrical signals, observed as areas of low voltage on the endocardium. This is caused by the build-up of collagen in the tissue \cite{jadidi2016ablation, angel2015diverse}. This allows for the assessment of cardiac fibrosis in living patients through the identification of these areas with reduced endocardial voltage, utilizing voltage mapping techniques. Originally, this method was used to pinpoint areas of low voltage in individuals with congenital heart disease \cite{de2001three}. Currently, voltage mapping is used in various clinical research studies. It serves as a tool to guide catheter ablation therapy (Section \ref{sec:ablation}), aiding in the enhancement of arrhythmia-free survival rates among patients with AFib.

\begin{figure}
    \centering
    \includegraphics[width=1\linewidth]{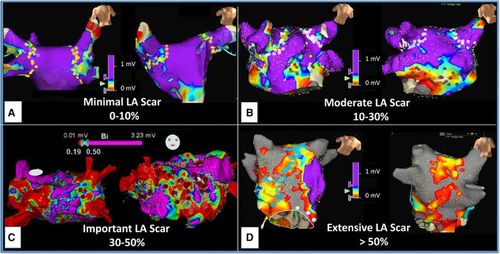}
    \caption{The approach of using low-voltage-guided ablation focuses on targeting specific areas within the left atrium exhibiting low-voltage patterns in patients with persistent AFib. This strategy involves conducting a detailed assessment with high-density multielectrode mapping, which maps more than 800 sites per atrium during active AFib episodes. Sites displaying a maximum bipolar voltage of less than 0.5 mV were identified as low-voltage areas. The comparative analysis of the LA low-voltage regions revealed their relative proportions as follows: 3\% (A), 24\% (B), 46\% (C), and 76\% (D) of the total LA surface area, excluding the pulmonary veins (PVs) \cite{jadidi2016ablation}.}
    \label{fig:lvz}
\end{figure}

Voltage mapping involves the detailed examination of the heart's potential substrates by capturing electrograms across various points on the atrial endocardium. During this process, a catheter is carefully moved across the endocardial surface, recording electrograms at each location in both unipolar and bipolar modes. Sites that show a voltage lower than a predetermined threshold (usually, this value is set as 0.5mV) are identified as part of low voltage zones (LVZ). By doing so, clinicians can differentiate between normal tissue and LVZs during sinus rhythm, indicative of fibrosis, which aids in creating detailed low voltage maps \cite{jadidi2016ablation}. Fig.~\ref{fig:lvz} shows the different  LVZ regions and how they determine the extent of the scars. 

\paragraph{At the Organ Level}

Atrial dilation is defined as the increase in the size of the atria, which is a notable part of the remodelling process. Atrial dilation is often linked with high heart rates and heart failure, observed across various studies and in AFib patients. The extent of dilation usually reflects the severity of the underlying condition and the impact of AFib. A significant echocardiographic measure in AFib patients is left atrial enlargement, which indicates eccentric hypertrophy due to increased pressure and volume \cite{jansen2020atrial}.

\subsubsection{Electrical  remodeling}

Electrical remodelling focuses on changes at the cellular and molecular levels, distinct from structural changes that occur across various biological dimensions \cite{allessie1977circus}. It primarily involves alterations in the action potential duration and the refractory period, with these modifications being reversible. This means the heart's electrical functions can return to normal shortly after a normal rhythm is re-established, showcasing the dynamic nature of electrical remodelling. The core elements of electrical remodelling include shifts in the movements of Na$^{+}$, K$^{+}$, and Ca$^{2+}$ ions. In electrical remodelling,  a decrease in L-type Ca$^{2+}$ current can be observed. This can lead to a shortened action potential duration due to an increased level of intracellular Ca$^{2+}$, making cells more prone to abnormal rhythms. Apart from that, changes in gap junctions can be observed, especially in the distribution of connexins like Cx40, Cx43, and Cx45, which affect the heart's electrical conduction system, particularly in people with long-standing AFib.

\subsubsection{Contractile remodelling}

In addition to electrical remodelling, AFib can also lead to contractile remodelling, affecting the heart muscle's ability to contract. This is due to changes in how muscle cells handle $Ca^{2+}$, which is crucial for muscle contraction. Research in animals and humans has shown that AFib can reduce the heart's pumping efficiency, which can be somewhat improved with calcium-channel blockers or worsened with drugs that increase calcium levels in cells. This suggests improper $Ca^{2+}$ handling at high heart rates may cause these changes. AFib can also cause heart muscle cells to revert to a less mature state, with a structure more resistant to calcium-induced damage but with fewer contractile elements, contributing to the development of a specific type of heart muscle disease (atrial cardiomyopathy).

\subsection{Diagnostic and Imaging Techniques in Atrial Fibrillation Management}

This section explores various methods for managing Afib, with a primary focus on the use of ECG for detection. Additionally, it discusses the application of MRI in identifying scar regions for the treatment of Afib, along with other techniques for its identification and management.


\begin{figure}[!htp]
    \centering
    \includegraphics[width=0.6\textwidth]{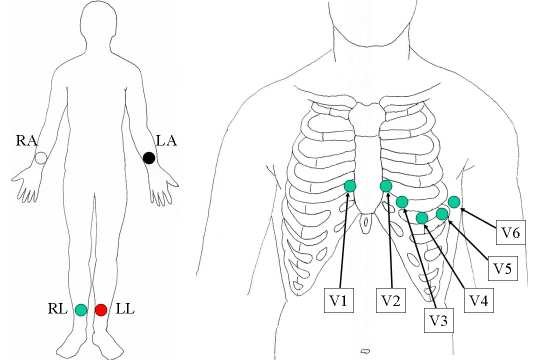}
    \caption{Electrode placement during 12 lead ECG. RL - Right Leg. LL- Left Leg, RA - Right Arm, LA - Left Arm, V1 - in the 4th intercostal space, right of the sternum, 
V2 - in the 4th intercostal space, left of the sternum, 
V3 - between V2 and V4, 
V4 - 5th intercostal space in the nipple line, 
V5 - between V4 and V6 and
V6 - in the midaxillary line on the same height as V4. Image is adapted from \cite{lemay2007data}. 
}
    \label{fig:ecg electroeds}
\end{figure}

\subsubsection{Electrocardiogram (ECG)}
\label{ecg-section}

\begin{wrapfigure}{r}{0.5\textwidth}
  \begin{center}
   \vspace{-5mm}
    \includegraphics[width=0.48\textwidth]{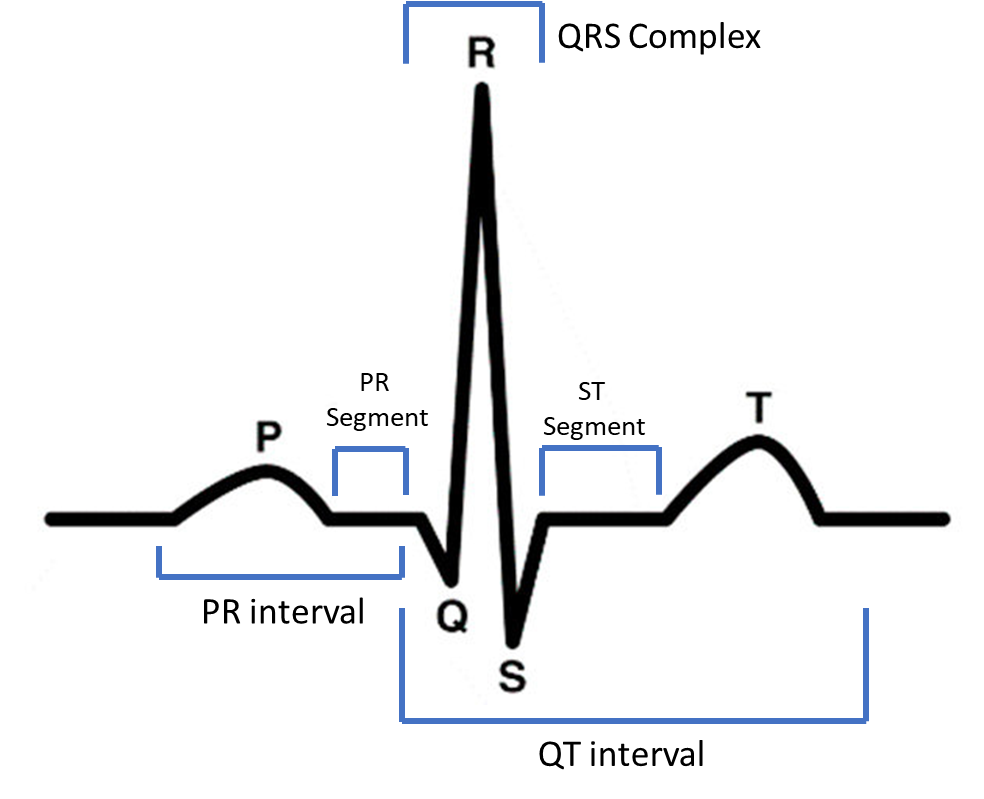}
     \vspace{-2mm}
  \end{center}
    \caption{Illustration of ECG segments and intervals. PR interval starts from the beginning of the P wave and ends with the beginning of the QRS complex. QT interval goes from the end of the PR interval until the end of the T wave.}
     \vspace{-5mm}
    \label{fig:ecg wave}
\end{wrapfigure}

An electrocardiogram, often abbreviated as ECG or EKG, is a medical diagnostic tool that is used to assess the electrical and muscular functions of the heart.This non-invasive procedure involves placing of small, flat electrodes on specific areas of the patient's chest and limbs. These electrodes detect the tiny electrical changes on the skin that arise during each heartbeat. The ECG works by capturing these electrical signals and translating them into a waveform representation, which is then displayed on a monitor or printed on paper. Each part of the ECG waveform provides valuable information about the electrical activity of different parts of the heart. For example, the P wave represents atrial depolarization, the QRS complex corresponds to ventricular depolarization, and the T wave indicates ventricular repolarization.

Usually, ECG is recorded on a standard graph paper, which consists of small and large squares.  The X-axis corresponds to the time, and the y-axis corresponds to the amplitude of the signal. In the ECG paper, the smallest square corresponds to 40ms, and the large square corresponds to 200ms. Hence, it takes five large squares for 1 second. In the clinical setting, 12-lead ECG is used for diagnostic purposes. A "lead" in ECG represents a graphical representation of the heart's electrical activity derived from data analysis from multiple ECG electrodes. In 12-lead ECG, 12 leads produce 12 separate graphs using only 10 electrodes. Six electrodes are placed in the chest, and four electrodes are placed in the limbs (right arm, left arm, right leg and left leg) (Fig.~\ref{fig:ecg electroeds})

\begin{figure}[!htp]
    \centering
    \includegraphics{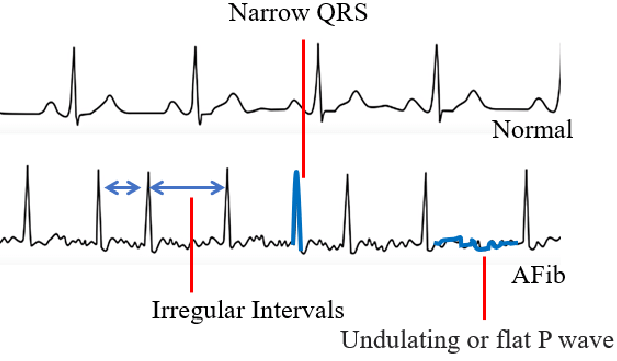}
    \caption{Difference between ECG during normal heart rhythm vs AFib. The upper ECG waveform shows the normal ECG, which consists of every feature, including P-wave, QRS complex and T-wave. In the below image, the absence of P-wave, irregular QRS complex and irregular RR interverals indicate the AFib. Image is adapted from \cite{bao2022paroxysmal}}
    \label{fig:afib ecg}
\end{figure}

The primary diagnostic instrument for AFib is the ECG. This condition is characterised by the absence of P-waves and irregular RR intervals on the ECG trace (as shown in Fig.~\ref{fig:ecg wave} and Fig.~\ref{fig:afib ecg}), distinguishing it from normal heart rhythm patterns \cite{stridh2008waveform}. According to the guidelines provided by the ESC in 2020,  \cite{hindricks20212020}, a minimum 30s recording with 12-lead or single lead ECG with heart rhythm without clear and preparing P waves and irregular RR intervals is considered diagnostic proof of clinical AFib.


\subsubsection{Magnetic Resonance Imaging (MRI)}
\label{mri}




Magnetic Resonance Imaging (MRI) is a non-invasive diagnostic technique that generates comprehensive internal body structure images encompassing organs, bones, muscles, and blood vessels. Unlike X-rays, which rely on ionising radiation, current MRI employs a powerful magnet coupled with radio waves to visualise internal structures. The core apparatus of MRI is a large cylindrical device, creating a potent magnetic field around the patient and utilising radio wave pulses for imaging. Various configurations of MRI machines exist, ranging from narrow tunnels to more open designs. Even though the fundamental physics behind MRI was introduced in 1946, it took more than 15 years to use it in clinical applications. Bloch et al. (1946) \cite{bloch1946nuclear} identified the magnetic properties of the atoms, and in 1971, Damadian et al. \cite{damadian1971tumor} used magnetic resonance for tumour detection. 

Currently, there is burgeoning interest in utilising Late Gadolinium-enhanced MRI (LGE-MRI) to assess atrial structure and aid in treatment planning for AFib. Moreover, the LA's diameter and volume, derived from 3D LGE-MRIs, have proven valuable in providing dependable data for determining treatment approaches \cite{njoku2018left}. This section focuses on the physics behind MRI and the usage of MRI in Afib.  

\paragraph{Physics Behind MRI}

The fundamental principle behind MRI lies in the response of water molecules within the body. According to the study conducted in \cite{mitchell1945chemical} regarding the chemical composition of the human body, the skin contains 65\% water, the skeleton 32\% water, the brain combines with the spinal cord and nerve trunks 73\% water, the liver 71\% water, the heart 73\% water and lungs 83\% water.  A water molecule contains two hydrogen atoms and one oxygen atom. Hydrogen atoms have only one electron (negatively charged) associated with the proton. It is positively charged and rotates its own axis. Since Hydrogen atoms have the highest ratio between magnetic dipole moment and angular moment (also known as gyromagnetic ratio), water molecules possess a slight magnetic field due to their inherent electric charges and continuous spin, known as nuclear spin. In a typical environment, these magnetic fields are randomly oriented and neutralise each other. 
\begin{wrapfigure}{r}{0.6\textwidth}
  \begin{center}
    \includegraphics[width=0.5\textwidth]{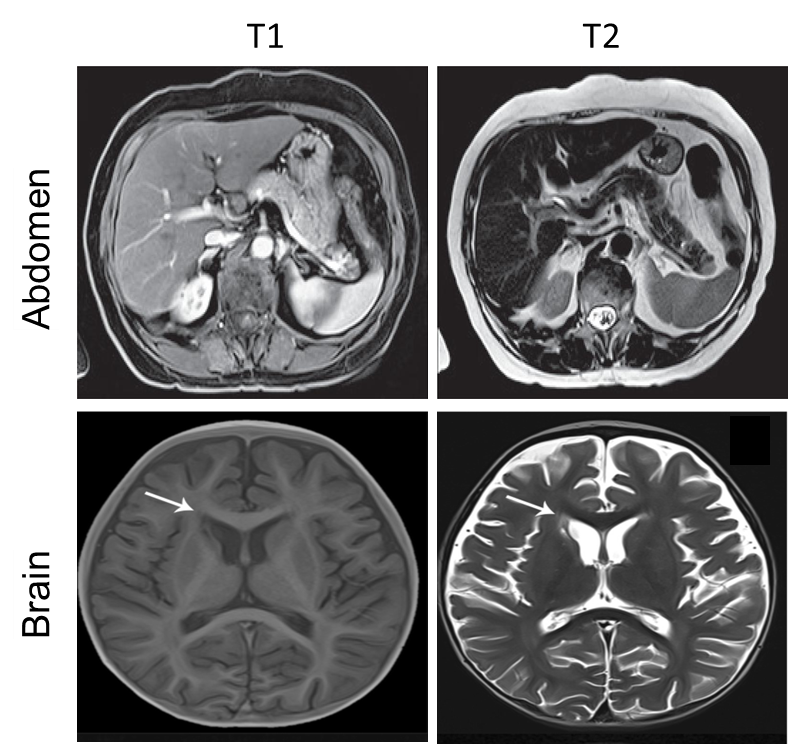}
  \end{center}
 \caption{Comparison between T1 and T2 weighted MRI of abdomen and brain \cite{zheng2021severe,santos2017intrapancreatic}}
   \vspace{-5mm}
\label{fig:t1vst2}
\end{wrapfigure}

However, when applying a strong magnetic field, these nuclei align along the direction of the field and spin coherently within their own axis. The imaging process begins when these aligned protons are disturbed by high-frequency impulses, resonating at a frequency matching their spin - the essence of the term “magnetic resonance imaging.” When the impulse ceases, the protons revert to their original alignment, emitting electromagnetic waves detectable by receivers. This reversion process, or “relaxation,” varies across different tissues, influencing the MRI signal’s strength and contributing to image contrast. This contrast is manipulated in MRI through “T1 weighted” and “T2 weighted” imaging, accentuating different tissue properties.

T1, also known as longitudinal relaxation time, can be defined as the time that takes the excited protons to come to their equilibrium state. It quantitatively measures the duration required for spinning protons to realign with the external magnetic field.  T2 or transverse relaxation time represents the time constant, which dictates the rate at which excited protons reach equilibrium or go out of phase with each other.  T1-weighted images highlight fat tissues, while T2 images highlight fluids in the human body. Visual comparison between T1 and T2 weighted images in different anatomical regions is shown in Fig.~\ref{fig:t1vst2}.

The MRI scanner’s architecture comprises several vital components. The primary element is the superconducting magnetic coil, generating a strong, constant magnetic field. Nested within are gradient coils that modify this field spatially, enabling precise localisation of the proton-induced electromagnetic signals. Closest to the patient are the body coils, producing the necessary high-frequency impulses and antennas for signal reception. MRI scanners are rated in tesla (T) to denote magnetic field strength, with clinical units ranging from 1.5T to 3T  and research units extending to 7T or more. Despite the high magnetic field strength, short-term exposure poses no significant risk to patients, although stringent safety measures regarding magnetic materials are imperative.

However, MRI is not universally applicable. Patients with magnetic implants, including certain pacemakers, cannot perform MRI scans due to the large magnetic field in the MRI machine. Additionally, the physical constraints of the machine limit its use for obese patients or those with claustrophobia. Its superior capacity for soft tissue differentiation is where MRI shines, making it preferable to computed tomography (CT) in many scenarios. Its non-reliance on ionising radiation further enhances its safety profile. Advancements in MRI technology have led to specialised applications like Magnetic Resonance Angiography (MRA), which assesses blood flow and detects vascular anomalies, and Functional Magnetic Resonance Imaging (fMRI). fMRI, particularly, has revolutionised brain mapping, enabling precise localisation of brain functions such as speech or memory, thus aiding in treatment planning for neurological disorders.

\paragraph{Late Gadolinium-enhanced MRI (LGE-MRI)}
\label{sec:lge_mri}

LGE imaging, a pivotal element in cardiovascular magnetic resonance (CMR) assessments, is conducted following the administration of gadolinium (Gd) contrast agents. This is a highly T1-sensitive imaging method.  Gd agents accumulate in areas with an expanded extracellular space, highlighting them in the images \cite{ylitalo2014late}. Often referred to as delayed enhancement imaging, LGE is instrumental in distinguishing between viable and non-viable heart muscle tissue across a broad spectrum of cardiac diseases. As the use of CMR becomes more widespread in the cardiovascular field, LGE imaging is increasingly being applied to patients with significant accompanying health issues, including acute heart failure and arrhythmia \cite{jenista2023revisiting}.

The underlying principle of this concept revolves around the delayed wash-in and wash-out in tissue with an increased proportion of extracellular space. Specifically, in acute myocardial infarction, this phenomenon is attributed to the combined effects of cellular necrosis, lysis, and edema. Conversely, in chronic infarcted tissue, fibrous scar tissue, which inherently possesses a larger extracellular space, serves as the primary factor \cite{doltra2013emerging}. This increased presence of the contrast agent, Gadolinium (Gd), is detectable through T1-weighted imaging techniques, typically within a 10-30 minute window post-contrast administration. 

Addressing the nature of Gadolinium, it is essential to note that in its free form, Gd is toxic. Consequently, contrast agents are formulated by binding Gd to a larger molecule, ensuring its safety during intravenous administration. The renal system predominantly handles the excretion of these agents, with a half-life of approximately two hours and near-total clearance from the bloodstream within 24 hours. Despite its longstanding reputation as a safe agent, recent insights have acknowledged the potential of Gd to induce nephrogenic systemic sclerosis, a rare yet severe condition \cite{doltra2013emerging}.

When considering clinical usage of LGE-MRIs in AFib treatment, a study conducted at the University of Utah has shown a correlation between increased fibrosis in the left atrial (LA) wall and outcomes in AFib ablation procedures \cite{oakes2009detection,akoum2013association}. In the study conducted by Chelu et al. (2018) \cite{chelu2018atrial}, the long-term efficacy of AFib ablation was examined, extending up to a five-year follow-up period. The inclusion criteria for participants in this analysis were twofold: firstly, they must have undergone an adequate quality LGE-MRI to evaluate LA fibrosis; secondly, they must have subsequently received AFib ablation treatment.

Current methodologies for examining atrial structures, particularly when utilising Late Gadolinium Enhancement Magnetic Resonance Imaging (LGE-MRI), predominantly depend on manual techniques. These techniques, while traditional, are noted for being excessively laborious and susceptible to inaccuracies. This situation underscores the necessity for a more efficient and automated approach to analyse atrial structures in three dimensions. With its rapid advancements, the burgeoning field of artificial intelligence (AI) presents a promising avenue. Consequently, many researchers are directing their focus towards leveraging deep learning algorithms. These algorithms can potentially expedite and enhance the accuracy of identifying atrial structures.

\subsubsection{Other Methods}

With the development of miniature devices and smartphones, numerous techniques for the early detection of AFib have emerged. A review conducted in 2019 ~\cite{li2019current} reported that over 100,000 mobile health apps and more than 400 wearable devices are available.  Those other methods are blood pressure cuffs initiated by patients, Photoplethysmogram (PPG) signal-based smart watches or wearables, patient-initiated ECG rhythm strips, wearable belts, ECG patches, etc. Although certain studies \cite{bumgarner2018smartwatch, wasserlauf2019smartwatch} evaluate the effectiveness of AFib detection through smartwatches, it's important for users to approach these results cautiously. This caution is advised because the apps and devices used may not have undergone clinical validation. Table \ref{tab:smart} shows the sensitivity and specificity values of various devices in AFib screening. However, it is notable that the evaluation of sensitivity and specificity across numerous studies often relied on small observational groups. These small cohorts significantly increase the potential for bias, particularly from signal selection. Additionally, the landscape of algorithms and technologies utilized in commercial devices continuously evolves, further complicating interpretation.

\begin{figure}[h]
    \centering
    \includegraphics[width=0.8\textwidth]{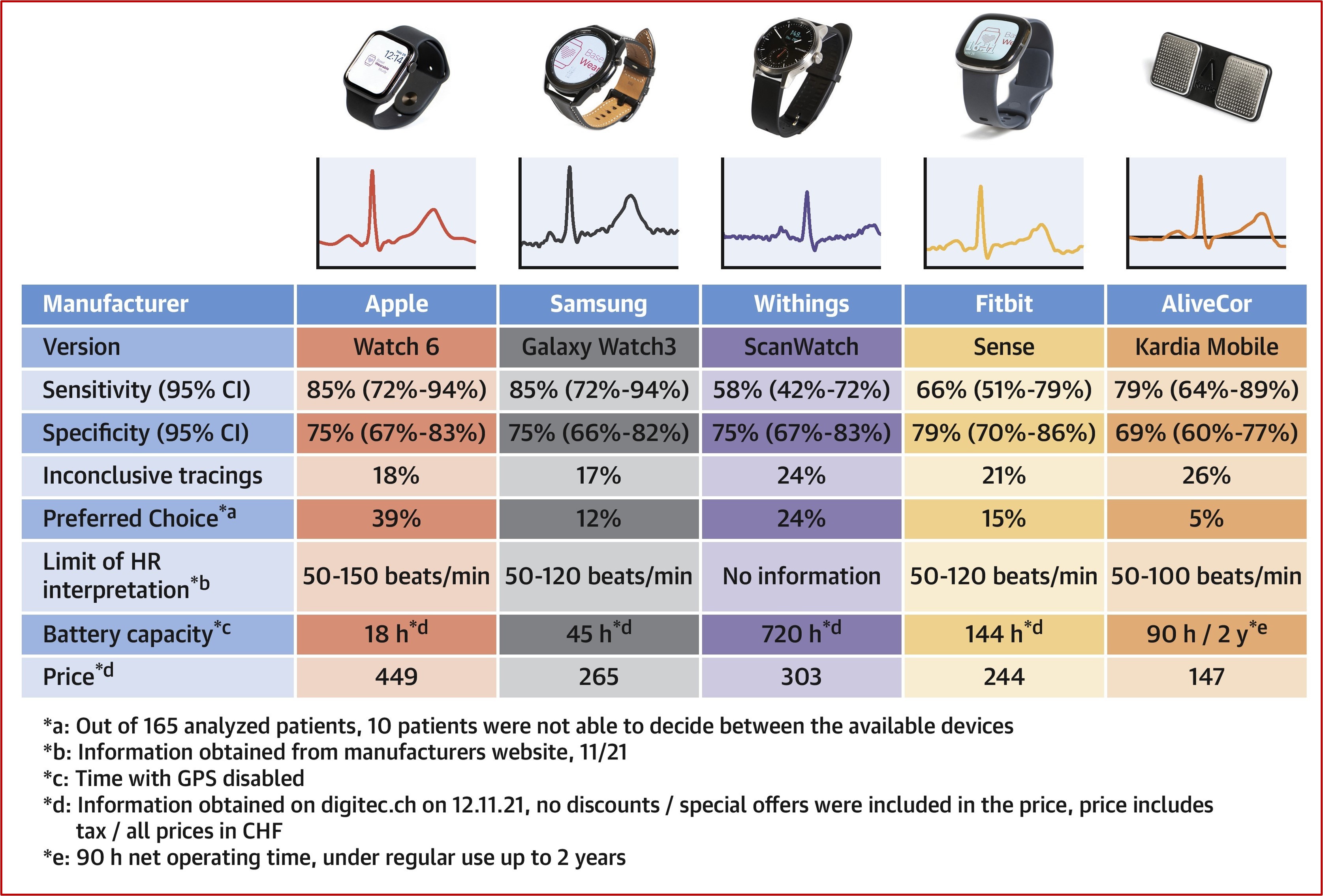}
    \caption[Comparison of analyzed wearable devices]{Comparison of analyzed wearable devices. CHF = Swiss Franc; GPS = Global Positioning System; HR = Heart Rate.}
    \label{fig:wearblae-Afib}
\end{figure}

The most exciting finding about additional screening methods came from the Apple Heart Study \cite{turakhia2019rationale} (AHS) and the Huawei Heart Study \cite{guo2019mobile} (HHS). The AHS included 419093 participants who signed up through a smartwatch app in the USA, with an average age of 40 years. Of these, 0.5\% were alerted to an irregular pulse rate, with younger participants under 40 years receiving notifications at a rate of 0.15\% and those over 65 at 3.2\%. Follow-up ECG patch monitoring for a week post-notification detected AFib in 34\% of those monitored. The HHS encompassed 187,912 individuals, predominantly male (86.7\%), with an average age of 35 years. Among those, around 0.23\% of participants received a notification for suspected AFib. Follow-up showed that 87\% of these cases were confirmed to have AFib, demonstrating a high accuracy (positive predictive value) of 91.6\% for detecting AFib using PPG signals, within a confidence interval of 91.5 to 91.8. In a subsequent study, Mannhart et al. ~\cite{mannhart2023clinical} evaluated the sensitivity and specificity of five smart devices in detecting atrial fibrillation using a sample of 165 patients, as depicted in Fig.~\ref{fig:wearblae-Afib}. The study found that among the devices tested, Apple and Samsung smart devices exhibited superior performance compared to the others.

While these additional screening methods help in preventing strokes, atrial remodelling, and morbidity and mortality related to AFib, as well as improving outcomes for conditions and diseases associated with AFib, incorrect interpretation of the signals and data can lead to anxiety from false positives, over-diagnosis, and over-treatment. This, in turn, might result in unnecessary invasive tests and treatments. Hence, a physician should confirm these results by using 12 lead ECG with more than 30s window.

\begin{table}
\caption{Sensitivity and specificity of various AFib
screening tools- All values are indicated as (\%) values and 12 lead ECG. The table is modified from \cite{mairesse2017screening}}
    \centering
    \begin{tabular}{llll}
    \toprule
         Methodology&  Study/Studies&  Sensitivity& Specificity\\ \midrule
         Pulse Taking&  \cite{harris2012can} &  87-97& 70-81\\
         Automated blood pressure measurements&  \cite{wiesel2004use, wiesel2009detection, stergiou2009diagnostic, willits2014watchbp}
 &  93-100& 86-92\\
         Single lead ECG screening&  \cite{desteghe2017performance, kaasenbrood2016yield, wiesel2013screening, jacobs2018cost}
&  94-98& 76-95\\
         Smartphone Apps&  \cite{yan2018contact, orchard2016screening, lowres2014feasibility, william2018assessing} 
 &  91.5-98.5& 91.4-100\\
         Watches&  \cite{nemati2016monitoring, tison2018passive, bumgarner2018smartwatch}
&  97-99& 83-94\\  \bottomrule
    \end{tabular}
    
    \label{tab:smart}
\end{table}



\subsection{Treatment for Atrial fibrillation}

The AFib treatment process is known as the ABC pathway \cite{hindricks20212020}, and it consists of three main parts. \textbf{A}- Anti-coagulation for stroke prevention, \textbf{B}- Better symptom control and \textbf{C}- Cardiovascular risk management. Treating complex arrhythmia like AFib presents challenges, partly due to the inadequate understanding of AFib mechanisms and the limitations of traditional mapping techniques in pinpointing arrhythmogenic substrates. This section explores the treatments for AFib in detail under the ABC pathway.

\subsubsection{Anti-coagulation for stroke prevention}

The focus here is on evaluating both the risk factors for stroke and the potential for bleeding. AFib quintuples the likelihood of experiencing a stroke, yet this risk varies based on the presence of specific factors that can increase the likelihood of a stroke. These risk factors are encapsulated within the CHA$_2$DS$_2$-VASc scoring system \cite{lip2010refining}, a clinical tool used to evaluate stroke risk. This system assigns points for the following conditions: Congestive Heart Failure, Hypertension, Age (75 years or older), Diabetes Mellitus, Prior Stroke or Transient Ischemic Attack, Vascular Disease, Age (65-74 years), and Sex category (Female). It's important to note that being female is considered an age-dependent modifier of stroke risk rather than a direct risk factor \cite{friberg2012assessment,tomasdottir2019risk, wu2020female}. Within this scoring system, a history of stroke and age of 75 years or older are each allocated two points, while all other factors are assigned one point each.

When starting antithrombotic treatment, it's important also to consider the likelihood of bleeding. Compared to placebo or no treatment, therapy with vitamin K antagonists, primarily warfarin, has been shown to decrease the risk of stroke by 64\% and reduce overall mortality by 26\% \cite{hart2007meta}. This is still widely used among patients with AFib globally. Moreover, in four key randomized controlled trials, the non-vitamin K antagonist oral anticoagulants, including apixaban, dabigatran, edoxaban, and rivaroxaban, demonstrated their ability to prevent stroke/systemic embolism just as effectively as warfarin. To minimize the risk of bleeding, it is crucial to maintain the quality of treatment with vitamin K antagonists and carefully choose the correct dosage of non-vitamin K antagonist oral anticoagulants.

\begin{figure}[!h]
    \centering
    \includegraphics[width=0.7\linewidth]{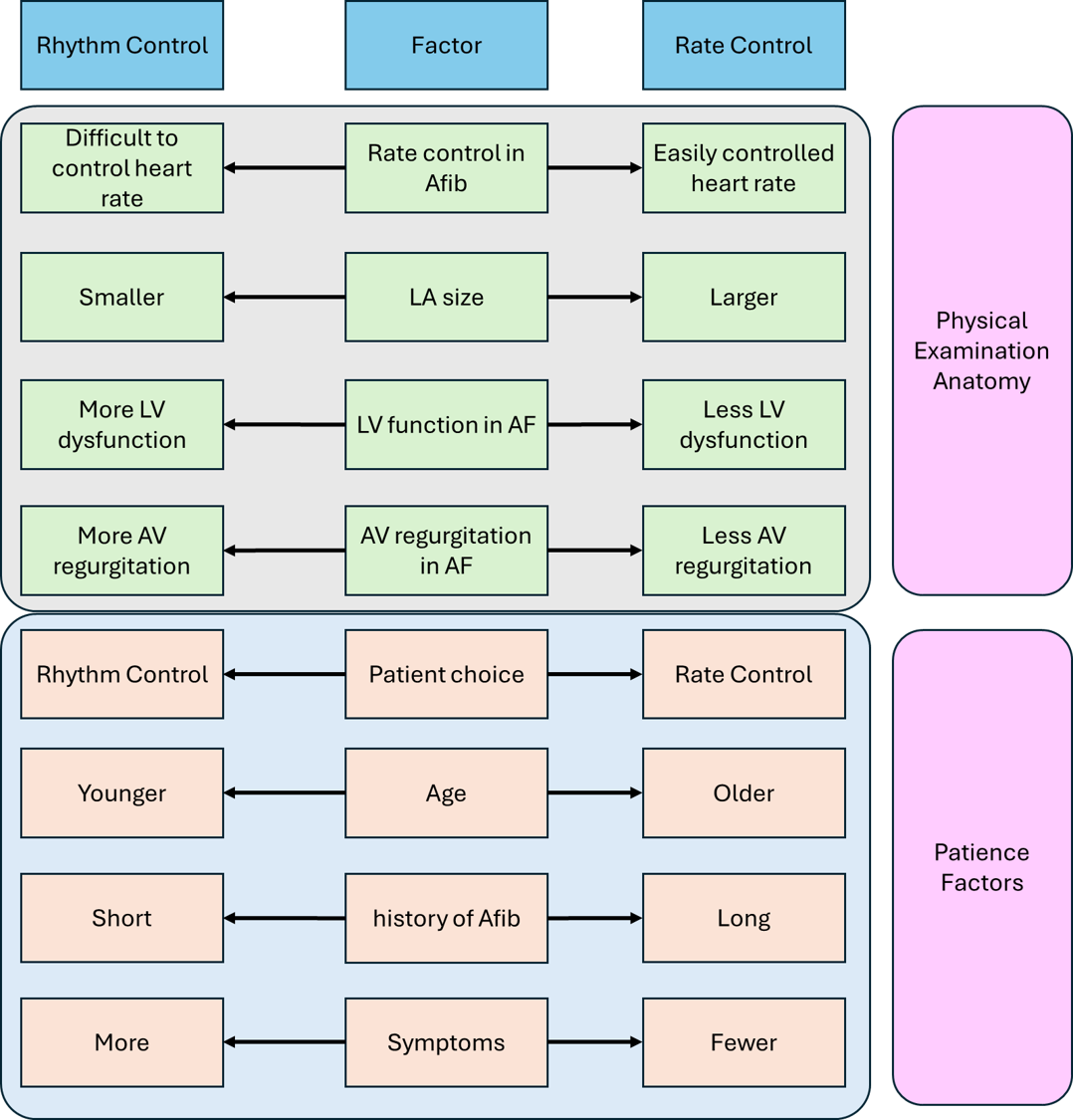}
    \caption{Considerations for choosing between rhythm and rate control strategies in patients with significant AFib burden. AV-atrioventricular; LA - left atrium; and LV - left ventricular. Figure is modified from \cite{joglar20242023}}
    \label{fig:rhythm_vs_rate}
\end{figure}

\subsubsection{Better symptom control}

Symptom control can be divided into two categories. 1) Rate control and 2) Rhythm control. Deciding between rhythm control and rate control is a critical step in the treatment pathway, and Fig.~\ref{fig:rhythm_vs_rate} outlines the factors to consider for patients with a substantial burden of AFib.

\vspace{1em}
\paragraph{Rate Control}

Rate control includes managing ventricular rate. In the initial stage, lenient rate control is applied, aiming for a target heart rate of less than 80 bpm at rest and less than 110 bpm during moderate physical activity. However, when symptoms are severe, a stricter rate control is applied with a target of less than 110 bpm. 

The most common strategy for rate control is the use of drugs. Beta-blockers (Atenolol, Propranolol, and Metoprolol), calcium channel blockers (Diltiazem and verapamil),  digitalis (Digoxin and Digitoxin) and Amiodarone are used as the rate control medications. Beta-blockers are frequently the primary choice for controlling heart rate, primarily due to their superior efficacy in achieving acute rate control. Calcium channel blockers should be avoided when the patient has heart failure with reduced ejection fraction \cite{joglar20242023}. Amiodarone may serve as a final option for managing heart rate in patients ineligible for non-pharmacological rate control methods, such as atrioventricular node ablation and pacing, especially when combination therapy fails to provide control. 

The pace-and-ablate strategy, which involves atrioventricular (AV) node ablation followed by the installation of a permanent pacemaker, offers an effective method for managing the heart rate and symptoms in patients with AFib  \cite{rijks2023left}. The pacemaker ensures that the heart beats at an appropriate rate, preventing it from beating too slowly. This approach is particularly useful when medications are ineffective in controlling the ventricular rate. Characterized by its straightforwardness, this procedure boasts a minimal risk of complications and a low mortality rate over the long term \cite{lim2007ablate}. Optimal results are often achieved by implanting the pacemaker a few weeks prior to performing the AV node ablation and setting the initial pacing rate between 70 and 90 beats per minute post-ablation. Notably, this treatment does not detrimentally affect left ventricular (LV) function and may even lead to improvements in left ventricular ejection fraction (LVEF) among certain patients \cite{hindricks20212020}. 

\vspace{1em}
\paragraph{Rhythm Control}

Therapies for rhythm control are designed to restore and maintain a normal, steady heart rhythm. This helps to avoid the return of the arrhythmia and prevents additional changes to the atrial structure(remodelling). Apart from that, compared to rate control, rhythm control shows lower AFib progression \cite{hindricks20212020}.

\vspace{1em}
\noindent \textbf{Cardioversion}

Cardioversion can be divided into two main parts: Pharmacological cardioversion and electrical cardioversion. Pharmacological cardioversion includes using medicines to restore the normal rhythm, while electrical cardioversion includes using a defibrillator to provide an electrical shock to restore the normal rhythm \cite{nusair2010electric}. Cardioversion is widely used if the patient is having AFib for the first time, and electrical cardioversion is the most used procedure compared to pharmacological cardioversion.  While electrical cardioversion tends to be more effective than its pharmacological counterpart, it requires the patient to undergo anaesthesia or sedation \cite{joglar20242023}. Additionally, the risk of blood clots and the need for anticoagulation measures are important considerations for both electrical and pharmacological cardioversion.

Skin burns can occur during electrical cardioversion. Flecainide, Propafenone and Amiodarone are widely used for pharmacological cardioversion, and Fecainide and Propafenone can be self-administered for selected patients as a "pill in the pocket" strategy. After undergoing cardioversion, it is important for patients to have a follow-up evaluation to ascertain if a different rhythm management strategy, such as AFib catheter ablation or a rate control method, should be considered in place of the existing treatment.

\vspace{1em}
\noindent \textbf{Ablation}
\label{sec:ablation}

\begin{figure}[!htp]
    \centering
    \includegraphics[width=0.7\textwidth]{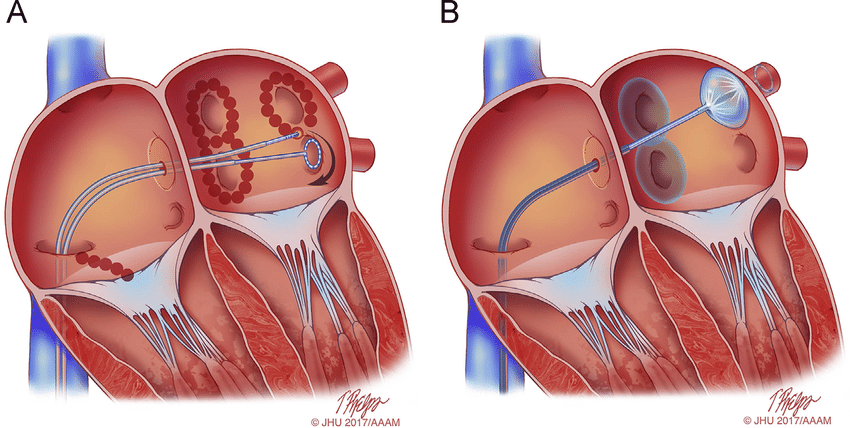}
    \caption{An illustration of catheter ablation.  \textbf{A:} illustrates the extensive area lesion set typically formed using RF energy, where ablation lesions follow a figure-eight pattern encircling the left and right pulmonary veins (PVs).  \textbf{B:} focuses on the cryoablation method. Here, ablation lesions are created around the right PVs, and the cryoablation catheter is situated in the left superior PV. Image is adapted from \cite{badhwar20172017}.}
    \label{fig:catheter ablation}
\end{figure}

Ablation is a therapeutic procedure which involves burning or freezing in some areas of the heart to disrupt erratic electrical signals. There are two main types of ablations: Maze procedure and catheter ablation.

 \underline{Maze procedure}

 \begin{figure}[!htp]
    \centering
    \includegraphics[width=1\linewidth]{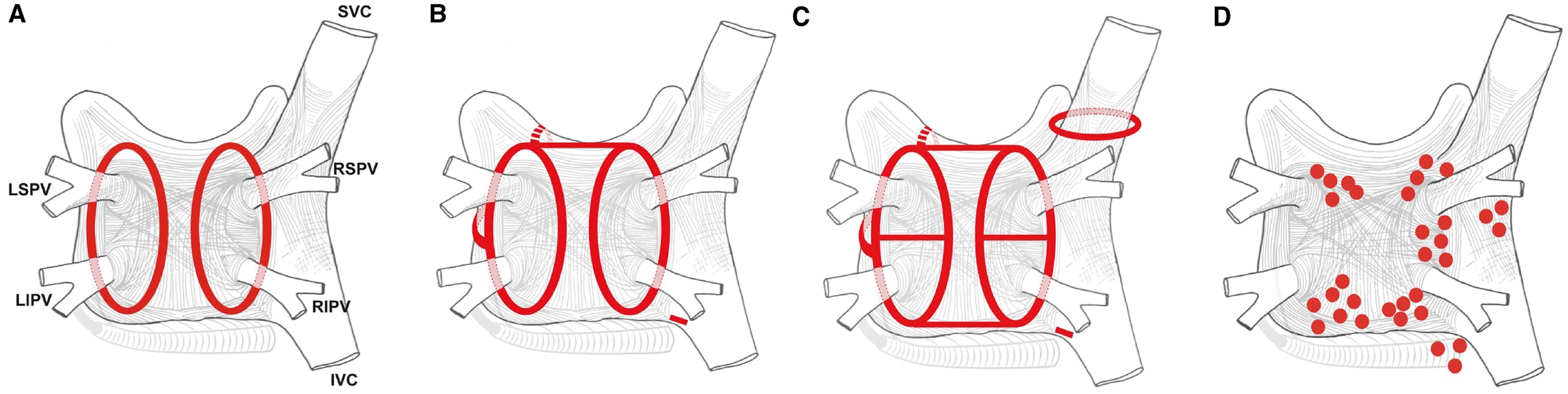}
    \caption{Overview of AFib Ablation Techniques: (A) Circumferential ablation around the pulmonary veins (PVs) aiming for PV isolation. (B) Common linear ablations include a roof line across PVs, a mitral isthmus line, and anterior linear lesions, with additional cavotricuspid isthmus ablation for certain atrial flutter cases. (C) Enhanced strategy showing figure of eight lesions around the PVs and additional lines for posterior left atrial wall isolation, including SVC isolation for focal firing. (D) Sites for ablation targeting rotational activities or complex fractionated atrial electrograms (CFAEs). Adapted from \cite{calkins20182017}}
    \label{fig:abaltaion_over}
\end{figure}
 
The Maze procedure is a bit complex surgical procedure which involves open-heart surgery. This was developed in 1987 by Cox et al.\cite{cox2000development}. It is often performed alongside treatments for other cardiac conditions, such as valve diseases or coronary artery disease. In this procedure, surgeons strategically create lesions on the heart's area that regulates the heartbeat. These lesions eventually form scar tissues, which block irregular electrical signals and thus help to normalize heart rhythms. The procedure may involve a combination of cryoablation (using freezing techniques) and radiofrequency ablation (using heat), depending on the recommendations of the medical team. With a success rate ranging between 80 to 90\%, the Maze procedure offers long-term relief from AFib. However, it's not uncommon for patients to experience skipped heartbeats or brief AFib episodes during the initial three months post-surgery, attributable to swelling in the atrial tissue. These symptoms are typically managed with medication and usually resolve as the heart heals.

 \underline{Catheter Ablation}

\begin{figure}
    \centering
    \includegraphics[width=0.6\textwidth]{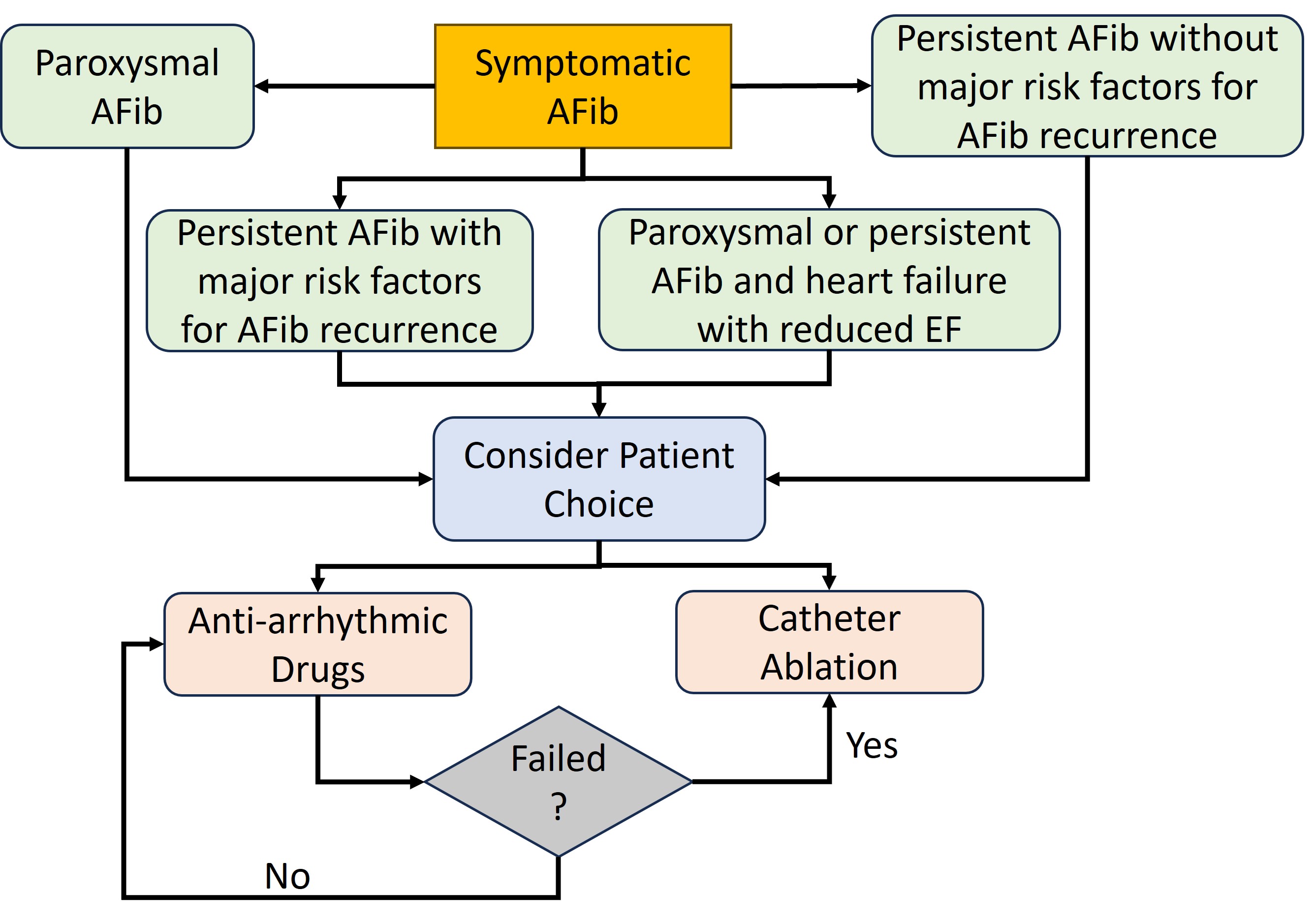}
  \caption{Flow chart of deciding needs of catheter ablation. Image is modified from \cite{hindricks20212020}. AFib = Atrial Fibrillation; EF- Ejection Fraction}
\label{fig:ablation-selectytion}
\end{figure}

Catheter ablation is a minimally invasive method compared to the Maze procedure. In this method, a thin small tube, referred to as a sheath, is inserted through the skin into a vein, creating a pathway. This is typically done via a vein or artery in the groin, although veins in the arm or neck can also be used.  Then, electrode catheters, which are thin, wired tubes, are guided through the sheath. These catheters are manoeuvred through the vein or artery to reach the heart, with X-ray imaging used for precise placement. After that, the catheter is employed to direct heat or cold energy to the targeted heart tissue, effectively eliminating the source of irregular rhythms. Finally, both the catheter and sheath are withdrawn from the vein or artery. An illustration of catheter ablation is shown in Fig.~\ref{fig:catheter ablation} and Fig.~\ref{fig:abaltaion_over}. Deciding process of the need for catheter ablation for a patient is shown in Fig.~\ref{fig:ablation-selectytion}.


In contrast to maze surgery, which aims to redirect erratic electrical impulses through the creation of scar tissue, catheter or cardiac ablation directly targets and destroys the tissue responsible for the arrhythmia. The foundation of catheter ablation lies in fully isolating the pulmonary veins through linear lesions around their entrance, achieved by either point-by-point radiofrequency ablation or single-shot ablation techniques. Data from prospective registries indicate that about 4 to 14\% of patients who undergo AFib catheter ablation face complications, with 2 to 3\% of these complications being potentially life-threatening \cite{szegedi2019repeat}. This underscores the necessity for ongoing monitoring post-procedure. 

\begin{figure}[ht]
    \centering
    \includegraphics[width=0.8\textwidth]{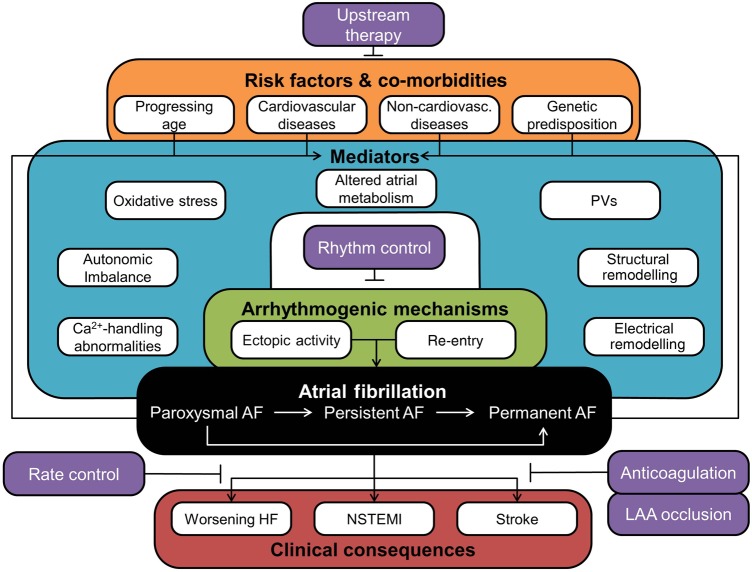}
\caption[Summary of the pathophysiology of AFib]{Summary of the pathophysiology of AFib. The black box highlights the progressive nature of AFib. The red box outlines the clinical consequences. The green box indicates that AFib is initiated and sustained by two primary arrhythmogenic mechanisms, which are influenced by various mediators (teal box). The orange box indicates several risk factors and comorbidities, while the purple boxes show the various therapeutic strategies to manage AFib. Key terms: HF (heart failure), LAA (left-atrial appendage), NSTEMI (non-ST segment elevation myocardial infarction).}
    \label{fig:afib_summary}
\end{figure}


\subsubsection{Cardiovascular risk management}

The "C" element in the ABC pathway focuses on recognizing and addressing concurrent diseases, cardiometabolic risk factors, and unhealthy lifestyle habits. The management of these risk factors and cardiovascular conditions plays a crucial role in enhancing stroke prevention efforts, lowering the burden of AFib, and alleviating symptom severity. Lifestyle modifications are essential components of this approach, encompassing strategies for obesity and weight management, moderation in alcohol and caffeine consumption, and encouraging physical activity. It is recommended that individuals engage in moderate-intensity exercise and maintain an active lifestyle to mitigate the risk of AFib development or recurrence, although it's advised to steer clear of extremely intense endurance activities. Key cardiovascular risk factors and comorbid conditions to be mindful of include hypertension, heart failure, coronary artery disease, diabetes mellitus, and sleep apnea. Patients are advised to manage these conditions proactively to minimize the likelihood of AFib recurrence.

In Figure~\ref{fig:afib_summary}, the pathophysiology of AFib is illustrated. The black box emphasizes the progressive nature of AFib. The red box highlights the clinical consequences of the condition, including worsening heart failure (HF), non-ST segment elevation myocardial infarction (NSTEMI) and stroke. The green box shows that AFib is initiated and sustained by two primary arrhythmogenic mechanisms: ectopic activity and re-entry, which are influenced by various mediators depicted in the teal box. The orange box indicates several risk factors and comorbidities associated with AFib, while the purple boxes present various therapeutic strategies to manage the condition.

\section{Deep learning}

\begin{figure}[h!]
  \centering
  \begin{subfigure}[b]{0.45\textwidth}
    \includegraphics[width=\textwidth]{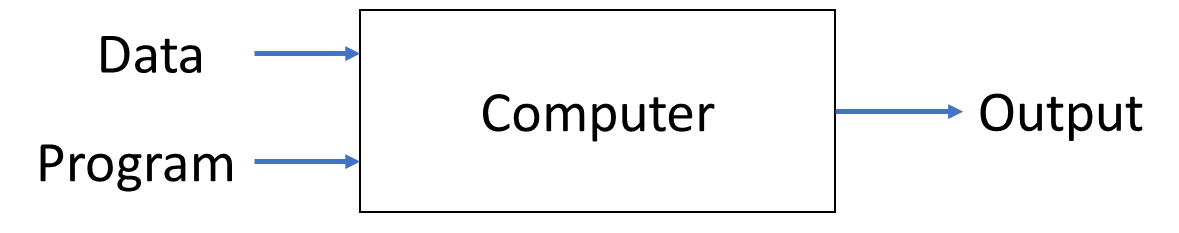}
    \caption{Traditional programming}
  \end{subfigure}
  \hfill
  \begin{subfigure}[b]{0.45\textwidth}
    \includegraphics[width=\textwidth]{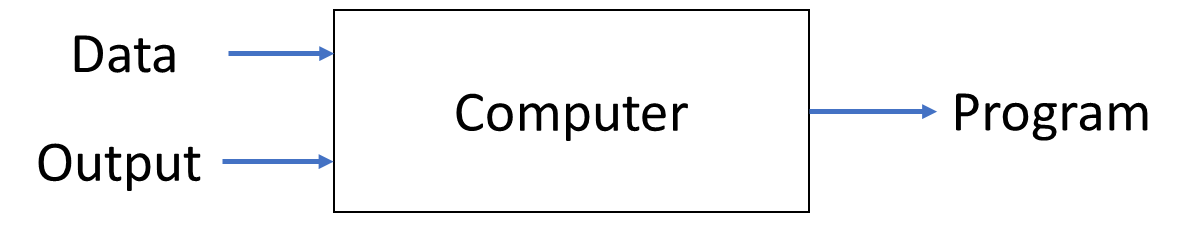}
    \caption{Machine learning}
  \end{subfigure}
    \caption{Difference between traditional programming and machine learning}
    \label{fig:mlvsprogram}
  \end{figure}

Artificial intelligence, or AI, has been a commonly heard word since the late 2000s. According to the Britannica Encyclopedia, AI is defined as the ability of a digital computer or computer-controlled robot to perform tasks commonly associated with intelligent beings \cite{bindushree2020artificial}. AI includes reasoning, learning, perception, problem-solving, language understanding, and more. Alan Turing made an early attempt at AI during World War II. However, these early methods are based on hard coding the knowledge rather than machine learning. Since hard-code knowledge-based methods are unsuitable for generalisation, researchers moved into a subset of AI known as machine learning. A comparison between traditional programming and machine learning is shown in Fig.~\ref{fig:mlvsprogram}. Machine learning models are tasked to learn from the raw data on their own and make decisions based on it. Machine learning uses statistical methods to enable machines to improve at tasks with experience. Among many ML techniques, deep learning has taken the attention of the AI community.  Deep learning models are structured in layers, creating an "artificial neural network" that can learn and make intelligent decisions independently, and deep learning requires a large amount of data to perform well. Deep learning has become a trend in areas like medicine, finance and physics. It is currently used for advanced applications like autonomous vehicles, facial recognition systems, and language translation services. Improvement of hardware processing units such as graphic processing units (GPUs) and the development of novel databases enable the rapid development of deep learning.

\subsection{Convolutional Neural Network}

Deep learning, a subset of machine learning in AI, is particularly adept at processing and analysing large datasets, which is instrumental in medical imaging.
\begin{wrapfigure}{r}{0.5\textwidth}
  \begin{center}
   \vspace{-5mm}
    \includegraphics[width=0.48\textwidth]{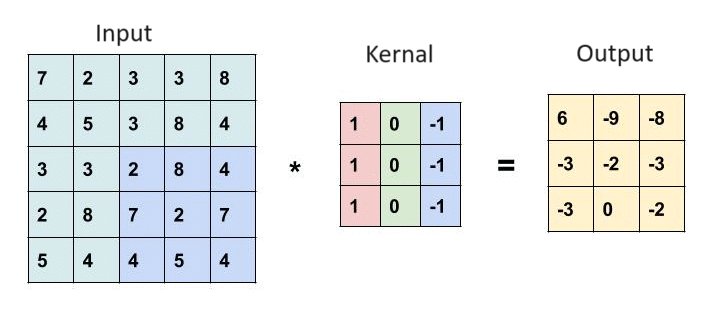}
  \end{center}
    \caption{Overview of the convolution}
     \vspace{-5mm}
    \label{fig:convo}
\end{wrapfigure}
By employing deep learning techniques, it is possible to automate the analysis of complex atrial structures, reducing the time and potential errors associated with manual analysis.The use of convolutional neural networks (CNNs) \cite{lecun2015deep}, a type of deep learning model (artificial neural network) designed explicitly for processing images, could be particularly beneficial in this context. CNNs can analyse the intricate patterns in LGE-MRIs, enabling precise identification and characterisations. 

CNNs are used to adeptly process data, such as images and videos, with a grid-like topology. A fundamental distinction between CNNs and traditional ANNs lies in incorporating convolutional layers within CNNs. These layers are instrumental in distilling information from the grid of pixels that constitute images, with each pixel value quantifying the intensity at a specific location, typically ranging from 0 to 255. In contrast to ANNs, CNNs exploit spatial hierarchies, allowing them to efficiently encode the locational data by considering the interdependence of adjacent neurons. This structural advantage significantly diminishes computational complexity, rendering CNNs more efficacious for tasks involving visual inputs.

The convolutional (Conv) layer is the fundamental block of the CNNs. The main operation in the conv layers is the multiplication between (dot product) two matrices. In this operation, a small matrix known as a filter moves across the input image or feature map of the last layer step by step, which is usually known as strides. Then, it performs element-wise multiplication with the part of the image it covers at each step. The sum of these products is then used to form a feature map (Fig.~\ref{fig:convo}). This process allows the network to learn specific features from the input data, such as edges, textures, or more complex patterns in deeper layers.


\begin{wrapfigure}{r}{0.5\textwidth}
  \begin{center}
   \vspace{-5mm}
    \includegraphics[width=0.48\textwidth]{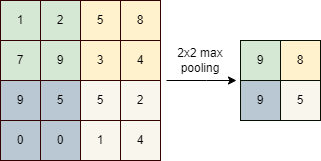}
  \end{center}
 \caption{2x2 Max Pooling }
  \vspace{-5mm}
    \label{fig:pool}
\end{wrapfigure}

The weights in the filter are learnable parameters. During training, the network adjusts these weights using backpropagation, allowing the network to focus on the most relevant features for the task at hand. Each unit in a convolutional layer is only connected to a small region of the input (known as the local receptive field), which makes convolutional layers spatially invariant and reduces the number of parameters compared to a fully connected layer. This local connectivity ensures that the learned features are spatially hierarchical, meaning lower layers learn simple features (like edges), and higher layers combine these to learn more complex features (like shapes or objects).

Often, a convolutional layer is followed by a pooling layer, which reduces the spatial dimensions (width and height) of the input volume for the next convolutional layer, leading to a reduction in the number of parameters and computations in the network. There are several pooling functions, such as max pooling and average pooling. Fig.~\ref{fig:pool} illustrates the 2x2 max pooling operation. 


In contemporary research, the application of CNNs is predominantly divided into three principal domains: classification \cite{he2016deep, simonyan2014very}, detection \cite{lin2017focal,redmon2016you} and segmentation \cite{ronneberger2015u}.

\paragraph{Classification}

Image classification is defined as the process of categorizing images into preset categories. Since CNNs can learn features without human intervention, it has been a hot topic in the deep learning community. Some popular classification architectures are VGG \cite{simonyan2014very}, ResNet \cite{he2016deep}, DenseNet \cite{huang2017densely}, SqueezeNet \cite{iandola2016squeezenet}, MobileNet \cite{howard2017mobilenets}, and ConvNext \cite{liu2022convnet} are some popular classification architectures.

\paragraph{Detection}

Detection takes the next step of the classification task. It involves localizing the object in images. In a nutshell, it means drawing a bounding box around the object. CNN-based object detection models can be divided into two main parts: one-stage networks and two-stage networks. Two-stage networks first extract the regions of interest (ROI). Then, those ROIs are used to detect the objects. Examples are R-CNN  \cite{girshick2014rich}, Fast R-CNN \cite{girshick2015fast} and Faster R-CNN \cite{ren2015faster}. In one-stage networks, ROI extraction is removed, and the object is detected directly. YOLO \cite{redmon2016you} can be taken as single-stage detectors.

\paragraph{Segmentation}

Image segmentation in computer vision can be conceptualised as a task involving categorising individual pixels using distinct semantic labels, a process known as semantic segmentation, or the boundary of specific objects, referred to as instance segmentation. Alternatively, it can encompass a combination of both these approaches, which is termed panoptic segmentation \cite{minaee2021image}. 

Semantic segmentation focuses on assigning a specific label from a predetermined set of categories, such as human, car, tree, or sky, to each pixel in an image. This task goes beyond general image classification, typically assigning a single label to an entire image, making semantic segmentation a more intricate and detailed process. The objective here is to gain a comprehensive understanding of the image at a pixel level instead of just a high-level overview.

Expanding upon semantic segmentation, instance segmentation delves deeper by categorising each pixel and differentiating and outlining each object within an image. For instance, in a scene with multiple people, instance segmentation would identify and delineate each person separately. This approach is particularly significant in applications where identifying and analysing individual items or entities within a group are crucial, such as in medical imaging, autonomous vehicles, and object tracking in video sequences.

The complexity of instance segmentation lies in its requirement to maintain the categorical information provided by semantic segmentation while segregating individual entities within the same category. This necessitates sophisticated algorithms capable of discerning subtle differences between objects of the same class and accurately delineating their boundaries. Thus, instance segmentation represents a more advanced and nuanced form of image analysis, providing a richer and more detailed interpretation of visual data than what is achievable through semantic segmentation alone. Mask R-CNN \cite{he2017mask} , Fully Convolutional Networks \cite{long2015fully}, U-Net\cite{ronneberger2015u}, V-Net \cite{milletari2016v}, nnU-Net \cite{isensee2021nnu}
, SegFormer \cite{xie2021segformer} , UniverSeg \cite{butoi2023universeg} and  Segment Anything (SAM) \cite{kirillov2023segment} are some of the widely used segmentation models.  There are many studies~\cite{xu2024dynamic,gunawardhana2024good, schilling2024assessment} that specifically focus on segmenting cardiac structures due to the specific nature of the heart (intensities,  different tissues in the heart, complex anatomy and dynamic nature)

\begin{figure}[!htp]
    \centering
    \includegraphics[width=\linewidth]{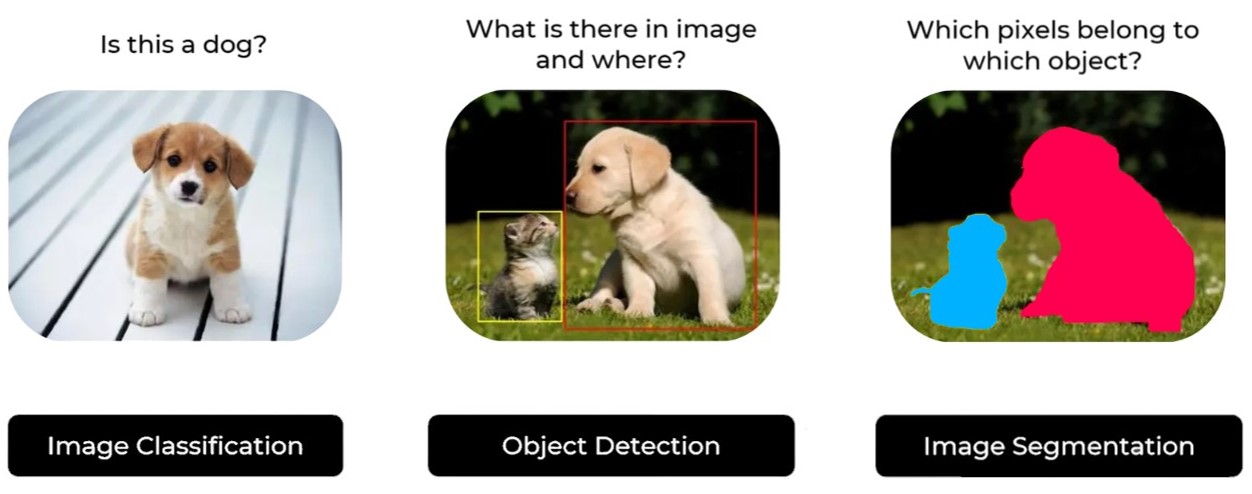}
    \caption{Difference between image classification, detection and segmentation}
    \label{fig:classificaiton_detection_segmentation}
\end{figure}

\begin{figure}[!htp]
    \centering
    \includegraphics[width=0.8\linewidth]{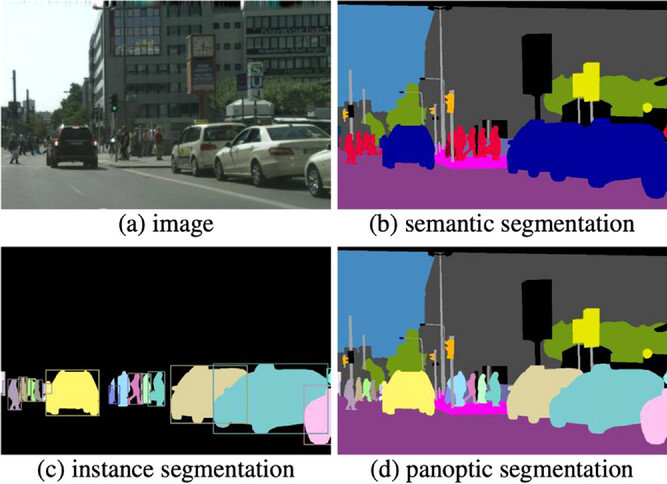}
    \caption{Difference between semantic segmentation, instance segmentation and panoptic segmentation}
    \label{fig:segment_methods}
\end{figure}

Fig.~\ref{fig:classificaiton_detection_segmentation} illustrates the distinctions among classification, detection, and segmentation while Fig.~\ref{fig:segment_methods} compares semantic segmentation, instance segmentation, and panoptic segmentation, highlighting their differences \cite{chuang2023deep}.

\subsection{Making deep learning model robust}

A deep learning model's ability to generalize effectively to unseen data largely depends on its capacity to learn the underlying patterns and information within the data. Enhancing the robustness of deep learning models is crucial for their performance, and several methodologies exist to achieve this. This section explores these techniques in detail, examining their roles in improving the model's ability to interpret and adapt to new, unseen datasets.

\subsubsection{Data Augmentation}

When applying data augmentation techniques, it is essential to ensure that relevant information is not removed from the images. For example, during image cropping, the focus should be on not eliminating regions of interest. Apart from that, depending on the application, the selection of appropriate data augmentation techniques is crucial. For instance, flipping chest X-ray images can result in images where the heart appears on the right side of the body, which could be misinterpreted as dextrocardia.

\subsubsection{Out of Distribution training}

Out-of-distribution (OOD) training is defined as training the model not only on a standard dataset but also exposed to data that is not representative of the distribution of the training dataset. The aim of OOD training is to improve the model's ability to generalize and perform reliably when it encounters data that deviates from the patterns and characteristics it was trained on. Exposing the model to a wide variety of data that may not be directly related to the task at hand helps the model to learn more generalized features. By training on OOD data, models are better equipped to handle unexpected inputs and scenarios in real-world applications, reducing the likelihood of failure or significant performance drops. OOD training is particularly important in safety-critical applications like autonomous driving, medical diagnosis, and financial forecasting, where encountering unexpected situations is expected and the cost of errors can be high. By training models to handle a broader range of inputs, OOD training seeks to enhance the reliability and robustness of machine learning systems.

\subsubsection{Regularization}

Regularization is a fundamental technique in machine learning, particularly in the context of deep learning, that aims to prevent overfitting. Overfitting occurs when a model learns the training data too well, including its noise and outliers, resulting in poor generalization to new, unseen data. Regularization techniques modify the learning process to encourage the model to be simpler or more general, which can improve its performance on unseen data. 

L1 regularization, L2 regularization and dropout are the most famous regularization methods in deep learning. 

\textbf{L1 and L2 Regularization}:  These techniques add a penalty term to the loss function based on the weights of the model. L1 regularization (Lasso) adds a penalty equal to the absolute value of the weights, promoting sparsity in the model weights (Eq. \ref{eq:l1}). L2 regularization (Ridge) adds a penalty equal to the square of the weights, encouraging smaller weights overall (Eq. \ref{eq:l2}).

\begin{equation}
\label{eq:l1}
   New\;loss = Loss function + \lambda \sum_{i=1}^{n} |w_i|
\end{equation}
\vspace{-0.5em}
\begin{equation}
\label{eq:l2}
   New\;loss =Loss function + \lambda \sum_{i=1}^{n} w_i^2
\end{equation}

\textbf{Dropout}: This involves randomly setting the a fraction of weight units to 0 at each update during training time, which prevents units from co-adapting too much. Dropout forces the network to learn redundant representations that are robust to the loss of any individual piece of information.

\subsection{Performance Evaluation Matrices}



Performance evaluation is a crucial aspect of deep learning, focusing on key metrics. Before delving into these metrics, it's important to define some terminologies: True Positives (TP) are correctly predicted positives, True Negatives (TN) are correctly predicted negatives, False Positives (FP) are incorrectly predicted positives, and False Negatives (FN) are incorrectly predicted negatives.

\subsubsection{Accuracy}

Accuracy is the ratio between number of correct predictions and total number of inputs.  However, the accuracy value can be misleading when the dataset is imbalanced. For example, assume in binary classification, out of 100 samples, 90 of them belong to one class. If the model predicts that category all the time, accuracy would be 90\%. To avoid that, the following matrices can be used.


\subsubsection{Precision} 

This is also defined as the Positive Predictive Value, this metric is the ratio of correctly predicted positive observations to the total predicted positive observations. High precision relates to a low rate of false positives.


\subsubsection{Recall (Sensitivity)}

This metric is the ratio of correctly predicted positive observations to all observations in the actual class. It is also known as Sensitivity or the True Positive Rate.

\begin{wrapfigure}{r}{0.5\textwidth}
  \begin{center}
    \vspace{-5mm}
    \includegraphics[width=0.48\textwidth]{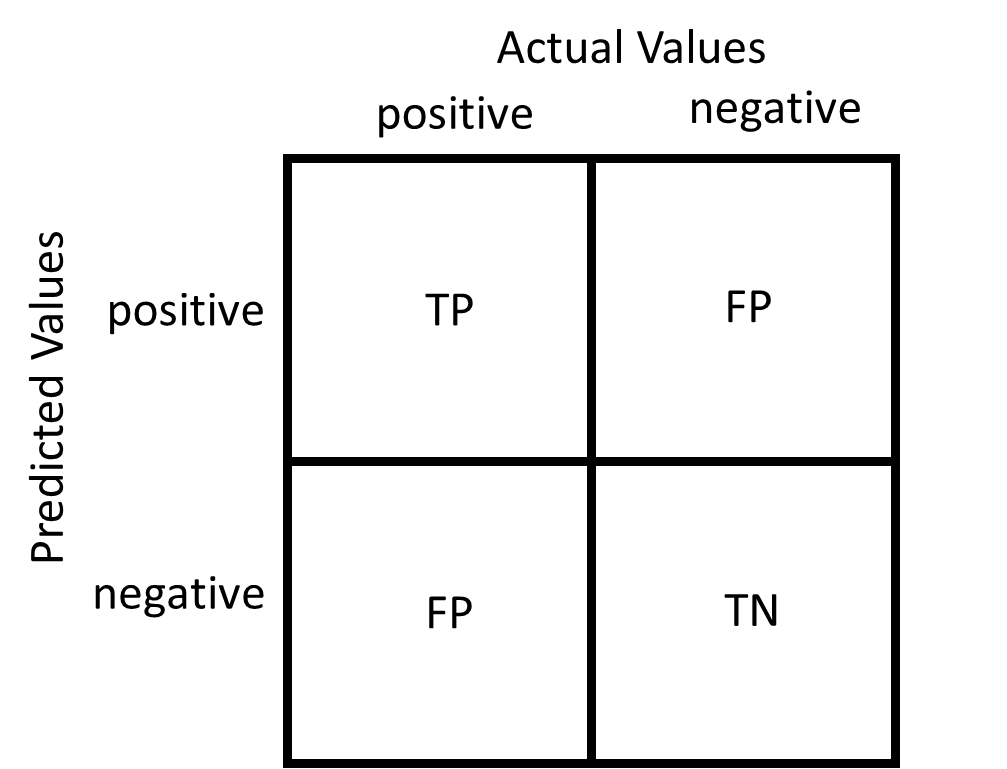}
  \end{center}
    \caption{Confusion Matrix}
     \vspace{-5mm}
    \label{fig:confmatrix}
\end{wrapfigure}

\subsubsection{F1 Score}

The F1 Score is the weighted average of Precision and Recall. Therefore, this score takes both false positives and false negatives into account. It is particularly useful when the class distribution is uneven.



\subsubsection{Confusion Matrix}

A confusion matrix (Fig.~\ref{fig:confmatrix}) is a table that is often used to describe the performance of a classification model on a set of test data for which the true values are known. It allows the visualization of the performance of an algorithm.

\subsubsection{AUC-ROC}

The Area Under the Receiver Operating Characteristic (ROC) curve is a performance measurement for classification problems at various threshold settings. ROC is a probability curve, and AUC represents the degree or measure of separability. It tells how much the model is capable of distinguishing between classes.


\begin{wraptable}{r}{8cm}
\centering
\renewcommand{\arraystretch}{2} 
 \vspace{-7mm}
\caption{Definitions of the performance metrics}

\label{tab:metrics}
\begin{tabular}{|l|c|}
\hline
\textbf{Metric} & \textbf{Definition} \\ \hline
Accuracy & \large$\frac{TP + TN}{TP + TN + FP + FN}$ \\ \hline
Precision & \large$\frac{TP}{TP + FP}$ \\ \hline
Recall & \large$\frac{TP}{TP + FN}$ \\ \hline
F1 Score & \large$2 \times \frac{Precision \times Recall}{Precision + Recall}$ \\ \hline
Dice Score & \large$\frac{2 \times |X \cap Y|}{|X| + |Y|}$ \\ \hline
Jaccard Index & \large$\frac{|X \cap Y|}{|X \cup Y|}$ \\ \hline
\end{tabular}
\end{wraptable}
\subsubsection{Dice Score}

The dice score measures the similarity between the predicted values and the ground truth value (in segmentation, those are the masks). It is defined as two times the intersection between predicted and ground truth masks divided by the sum of the areas. When the ground truth and predicted masks are the same, it achieves the highest value of one, and if there is no overlap, the dice score will be zero. Even though dice score calculation is simple, the main drawback is it is affected by the size of the comparing regions. 



\subsubsection{Jaccard Index}
The Jaccard index calculates the intersection between ground truth and predicted masks divided by the union between two masks. Similar to the dice score, the Jaccard index varies from zero to one. 

The relationship between the dice score and the Jaccard score can be shown as follows \cite{eelbode2020optimization}. 

\begin{equation}
    \text{Jaccard Score} \ = \ \frac{\text{Dice Score}}{2 \ - \ \text{Dice Score}}
\end{equation}

or 

\begin{equation}
     \text{Dice Score} \ = \ \frac{2 \ \times \ \text{Jaccard Score}}{\text{Jaccard Score} \ + \ 1}
\end{equation}

\subsubsection{Hausdorff Distance }

As the name suggests, Hausdorff distance measures the distance between two subsets. In image segmentation, it measures the boundary points of segmented regions. A smaller Hausdorff Distance indicates a closer match between the segmented boundary and the true boundary, implying higher accuracy of the segmentation algorithm. One characteristic of the Hausdorff Distance is its sensitivity to outliers. Since it is based on the maximum distance between boundary points, even a single pair of points that are far apart can significantly increase the Hausdorff Distance, even if the rest of the points are closely aligned. This can sometimes lead to misleading results, especially in cases where a few outlier points do not accurately reflect the overall quality of the segmentation. The equations for the performance metrics are summarized in Table~\ref{tab:metrics}.


\subsection{Challenges in deep learning}

Despite the widespread adoption and significant progress in deep learning, several prevailing challenges persist in this domain, which continues to be the focus of extensive research efforts.

\subsubsection{Data}

\paragraph{Data scarcity}

\begin{wrapfigure}{r}{0.5\textwidth}
  \begin{center}
    \includegraphics[width=0.48\textwidth]{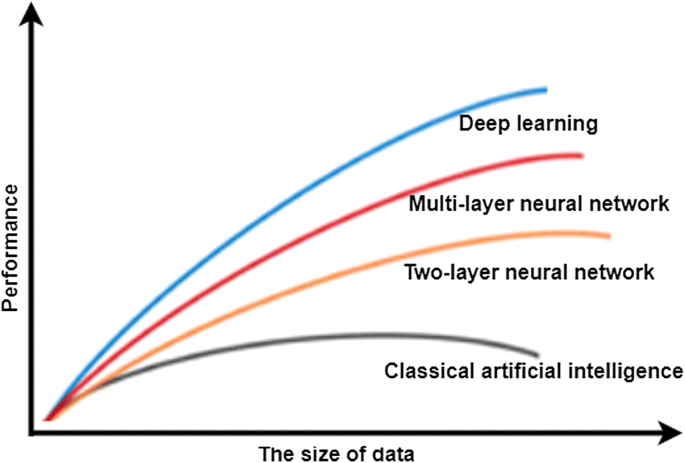}
  \end{center}
\caption{Performance of different machine learning approaches with respect to the size of the data\cite{serin2020review}}
 \vspace{-5mm}
    \label{fig:datavs model}
\end{wrapfigure}

Deep learning models are data-hungry. It needs extensive datasets to perform well. As illustrated in Fig.~\ref{fig:datavs model}, an increase in dataset size notably enhances the performance of deep learning models, often surpassing the performance of traditional machine learning methods. This is particularly evident in domains where large, diverse datasets are readily available, such as ImageNet for natural image processing \cite{deng2009imagenet}. However, the main challenge is when it comes to acquiring and preparing datasets for specialized fields such as medical imaging or financial analysis. These sectors not only require the collection of vast amounts of relevant raw data but also necessitate the involvement of experts for precise data labelling, making the process both labour-intensive and costly.

To address this bottleneck, several strategies have been employed to expand the size of datasets effectively. One of the easiest methods is data augmentation, which involves the application of various transformations to existing images, thereby artificially increasing the volume and diversity of the dataset. This technique is particularly useful in enhancing the robustness of deep learning models by presenting them with a broader range of data scenarios.

Another innovative approach is the utilization of synthetic data, a method gaining traction in the field of deep learning. Among the techniques for generating synthetic data, Generative Adversarial Networks (GANs) have emerged as a powerful tool \cite{goodfellow2020generative}. These networks are capable of producing highly realistic images, which can be used to supplement training datasets, especially in domains where data collection is challenging or restricted. The use of synthetic data is a rapidly evolving area of research, offering promising applications in real-world scenarios \cite{nikolenko2021synthetic}. The creation and integration of synthetic data pose unique challenges and opportunities. One of the primary challenges is ensuring the realism and variability of the generated data, which is crucial for the effective training of deep learning models. Additionally, the ethical and legal aspects of using synthetic data, particularly in sensitive fields like healthcare, require careful consideration.

Another approach is the use of transfer learning methods or unsupervised learning methods. Transfer learning, in particular, involves the application of knowledge acquired from one task (source task) to enhance performance in a different but related task (target task). This approach is exemplified in architectures like VGG and ResNet, which are initially trained on large-scale, annotated datasets like ImageNet. The learned weights from these models are often made publicly available, enabling researchers to utilize them as a starting point. By fine-tuning these pre-trained weights, it's possible to adapt the models to specific target tasks, often with less data and computational resources than training a model from scratch.

On the other hand, unsupervised learning-based methods capitalize on the use of unlabeled data. These methods focus on identifying inherent patterns and structures within the data without the guidance of explicit labels. This approach is particularly beneficial in scenarios where labelled data is scarce or expensive to obtain. By leveraging algorithms capable of extracting meaningful information from unlabeled datasets, researchers can uncover insights or prepare the data for further supervised learning tasks.



\paragraph{Bias in the datasets}

Biases in the data are one of the most challenging issues when developing state-of-the-art (SOTA) deep learning models because they are hard to identify. There are various types of biases, such as selection bias, framing bias, label bias, availability bias, exclusion bias, etc \cite{fabbrizzi2022survey}. Careful attention to data is required to alleviate the bias in the dataset. 

However, sometimes, even humans cannot detect biases in the dataset. One of the best examples of this is the study \cite{gichoya2022ai} conducted in 2022, which found that deep learning models can predict the race of human chest x-ray images without explicitly providing information about the race to the model. According to the authors of the study, some hidden information in the dataset led to this result. The authors of the study stressed the urgency of comprehending the underlying reasons enabling algorithms to possess such predictive capabilities. They highlighted the potential risks this could pose, especially to minority and underserved patient populations, if such biases are not adequately addressed in AI applications in healthcare, such as image-based diagnostic algorithms. This phenomenon challenges a prevailing assumption in the field of AI development: concealing demographic factors like race, gender, or socioeconomic status from AI models would prevent these models from discriminating based on these characteristics, thereby ensuring fairness. However, this study's findings indicate that such an approach is overly simplistic and insufficient for guaranteeing equity in AI and machine learning applications. 

Therefore, it becomes crucial for researchers and developers in AI and machine learning to not only focus on the overt aspects of dataset composition but also to delve into the more subtle, hidden patterns that might lead to biased outcomes. This necessitates a more nuanced and comprehensive approach to data analysis and model training, where fairness and equity become central tenets in the development of AI systems.

\subsubsection{Computational complexity}

Even with the availability of large datasets, most researchers face one of the fundamental problems while training deep learning models: computational complexity. The limitations of Central Processing Units (CPUs), primarily designed for sequential processing tasks, render them inadequate for handling the vast amounts of data typical in deep learning applications. This inadequacy has shifted the focus towards more specialized hardware solutions like Graphics Processing Units (GPUs) and Tensor Processing Units (TPUs), which are inherently designed for parallel processing. GPUs, originally tailored for rendering graphics, have architectures that allow for efficient handling of multiple operations simultaneously, making them well-suited for the matrix and vector operations central to deep learning. TPUs, on the other hand, are custom-designed by Google for TensorFlow, a popular deep-learning framework. They are optimized for both the training and inference phases of deep learning models, providing significant speedups and energy efficiency.

Despite their advantages, their high costs and substantial computational resource requirements often hinder the adoption of GPUs and TPUs in research settings. The financial burden is particularly acute for individual researchers or smaller institutions. Furthermore, the physical infrastructure needed to support these units, including advanced cooling systems and power supplies, adds to the complexity and expense. This challenge is exacerbated in the context of video-based deep-learning models. Such models demand not only high computational power due to the processing of sequential frames but also substantial memory capacity for storing large video datasets. Consequently, researchers working with video data frequently require multiple GPUs, further escalating the costs and resource requirements.

Researchers often resort to cloud-based solutions that offer GPU and TPU services to mitigate these challenges. While this approach can be cost-effective and scalable, it introduces dependencies on internet connectivity and potential concerns regarding data privacy and security. As the field of deep learning continues to evolve, there is a growing need for more accessible, efficient, and cost-effective computational solutions to enable broader research.

\subsubsection{Optimization}
When optimizing deep learning models, there are two main issues happening with respect to the gradient. 

\paragraph{Vanishing Gradient}

In deep learning, each layer's gradients are calculated based on the gradient of the layer ahead. As the gradient values are propagated back through the network, if these gradients are very small (less than 1), they tend to get smaller and smaller as they propagate through each layer. This diminishing effect results in an exponentially decreasing gradient as the backpropagation algorithm progresses through the layers. Consequently, neurons in the earlier layers learn very slowly, if at all, making it challenging to train deep networks effectively. This problem is particularly pronounced in networks with saturating activation functions, like the sigmoid or tanh functions.

 \paragraph{Exploding Gradient}
 
Conversely, the exploding gradient problem occurs when the gradients during backpropagation become very large. This typically happens in deep networks with many layers. As the error gradient is backpropagated, the gradients can accumulate and exponentially grow instead of diminishing, as in the vanishing gradient problem. The result is very large updates to the neural network weights during training, causing the learning process to diverge and the model to become unstable. This often leads to issues like numerical instability and the inability of the model to converge to a meaningful solution.

Various techniques have been developed to mitigate these problems. For the vanishing gradient problem, non-saturating activation functions like ReLU (Rectified Linear Unit) and its variants (e.g., Leaky ReLU) are often used. These functions help in maintaining a gradient that doesn't diminish as quickly as the traditional sigmoid or tanh functions. For the exploding gradient problem, gradient clipping is a common technique, where gradients are artificially capped during the backpropagation process to prevent them from becoming too large. Additionally, better weight initialization strategies (He initialization \cite{he2015delving}, Xavier initialization \cite{glorot2010understanding}) and batch normalization are also employed to alleviate these issues.

\section{Use of Artificial Intelligence for scar segmentation }

Integrating AI into the medical field could facilitate the development of predictive models that aid in the early diagnosis and treatment planning for atrial conditions. The ability of AI systems to learn from vast amounts of data can also contribute to a more personalised approach to patient care. By analysing historical data and identifying subtle patterns that might be overlooked in manual analysis, AI can provide insights crucial for patient-specific treatment strategies.  In the realm of scar segmentation from LGE MRI, researchers have identified four primary methodologies. The first approach involves the direct segmentation of scars from LGE MRI. This technique, however, presents difficulties due to the typically small volume of scars and the enhanced regions adjacent to them. The second strategy is to initially segment the left atrial wall and then classify each voxel within this segmentation as either normal wall tissue or scar tissue. This method, however, is hampered by the challenge of automatically segmenting the left atrial wall, a task complicated by its thin thickness. The third approach begins with the segmentation of the left atrial. Subsequently, the wall segmentation is generated by establishing a fixed distance from the left atrial endocardium. Utilizing this approximated wall segmentation, scars are then extracted from within it. The final approach diverges from the previous methodologies by bypassing the wall segmentation step entirely. Instead, this method involves directly projecting the segmented scar onto the LA surface for the purposes of scar quantification. This approach is particularly relevant in clinical studies, which tend to focus predominantly on the location and extent of scars, implying that the wall thickness can be disregarded in these analyses.

This section initially explores the intricacies of MRI data. Then, it is divided into two primary segments. The first part concentrates on existing competitions and the solutions proposed for current challenges in scar segmentation. The latter part delves into independent research that centres on scar segmentation.

\subsection{Deep learning for MRI data}

Natural images consist of 3 channels stacked on top of each other. These channels are known as RGB channels (Red, Green and Blue). Hence, most of the coloured natural images will be in the shape of Height x Width x 3 (HxWx3). However, MRI images are in HxWxDx4 format. Since MRI images are volumetric data, D represents the number of slices in the MRI scan. In a 3D MRI image, it is the number of individual 2D slices that are stacked together to form the complete 3D image. Number 4 illustrates the number of modalities/channels in the input data. For example, in brain imaging, common sequences are T1-weighted, T2-weighted, FLAIR, and Diffusion-weighted imaging. Each of these would be a separate channel. The number 4 suggests that four such channels or types of data are included in the dataset. Usually, MRI data is stored in DICOM (Digital Imaging and Communication in Medicine) or Nifti formats. 

\subsubsection{Obtaining scar data}

\begin{figure}
    \centering
    \includegraphics[width=\textwidth]{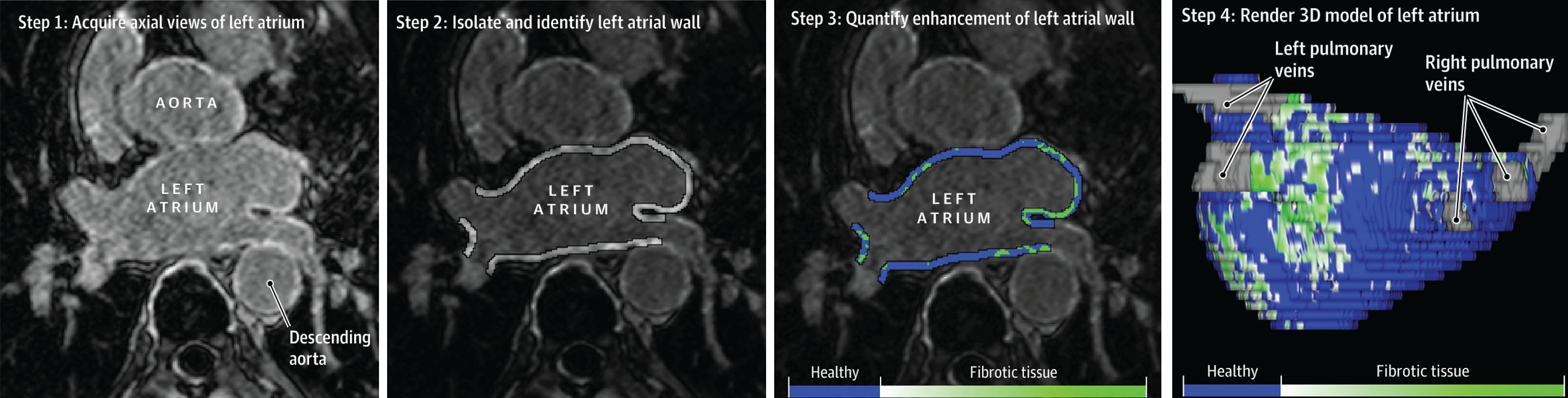}
    \caption[Atrial fibrosis quantification using LGE-MRIs]{Atrial fibrosis quantification using LGE-MRIs \cite{marrouche2014association} : (1) obtaining LGE-MRIs of the LA, (2) Segmentation of the LA wall from these images, (3) Quantification of fibrosis based on relative contrast enhancement and (4)  Projection of a colour-coded fibrosis map onto the LA wall.}
    \label{fig:decafff}
\end{figure}

As mentioned in Section \ref{sec:lge_mri}, LGE-MRI is a crucial tool for clinicians to detect fibrotic areas within the atrial wall. Quantifying fibrosis from LGE-MRIs involves four steps, as shown in Fig.~\ref{fig:decafff}. First, high-resolution 3D LGE-MRI images of the left atrium (LA) are acquired. Next, the LA wall is segmented from these images. Then, an appropriate thresholding method is used to identify the fibrotic regions. Finally, a color-coded fibrosis map is projected onto the LA wall. Several thresholding methods that can be used are discussed below.

\paragraph{Visual assessment} 

Recognized as the gold standard for evaluating ventricular fibrosis when histological validation is unavailable. This approach is inherently subjective, depending entirely on the observer's skill and experience. As a result, this manual technique faces challenges with consistency and the ability to be scaled up.

\paragraph{Standard Deviation Method}

This is a commonly used technique in medical imaging for identifying fibrotic areas in tissue. It involves setting a threshold based on a reference value derived from the intensity histogram of a relevant anatomical structure such as the blood pool or atrial wall. This reference value is adjusted by a manually selected multiple of the standard deviation to define the threshold. Mathematically, it is represented as:
\begin{equation}
\text{Th} = \text{Ref} + n \times \text{SD}
\end{equation}

where $\text{Th}$ is the threshold, $\text{Ref}$ the reference value, $\text{SD}$ the standard deviation, and $n$ the manually chosen multiplier. If a pixel value is greater than the obtained threshold, it will be considered fibrosis. \cite{mcgann2008new} used this threshold method for 386 patients and they were able to stratify patients into four stages of fibrosis, which proved critical for optimizing catheter ablation treatment outcomes.

\paragraph{Otsu Thresholding}

Otsu thresholding \cite{gunasekaran2020otsu} aims to minimize the intra-class intensity variance, effectively distinguishing between foreground and background. The method is formulated as:
\begin{equation}
s^2_w(t) = w_0(t) s_0^2(t) + w_1(t) s_1^2(t)
\end{equation}
where $t$ is the threshold separating the two classes $w_0$ and $w_1$ based on their variances, with $s_0^2$ and $s_1^2$ representing the respective class probabilities separated by $t$.

\paragraph{Fixed Width at Half Maximum (FWHM)}

The FWHM method, which requires no manual adjustment, is noted for its reproducibility in identifying fibrosis by setting a threshold at 50\% of the maximum signal intensity of the examined area. However, its effectiveness is limited in cases where scar tissue exhibits homogeneous gray values. FWHM has been primarily applied to post-ablation fibrosis detection in the LA and  is described mathematically as:
\begin{equation}
\text{Th} = \frac{\text{I}_{\text{max}} - \text{I}_{\text{min}}}{2}
\end{equation}
where $\text{Th}$ is the threshold, $\text{I}_{\text{max}}$ and $\text{I}_{\text{min}}$ are the maximum and minimum pixel intensities, respectively.

\paragraph{Image Intensity Ratio (IIR)}

This thresholding approach normalizes pixel values by the mean intensity of the LA blood pool to mitigate confounding factors such as contrast agent dosage, imaging protocols, or BMI. The IIR method is defined as:
\begin{equation}
\text{Th} =  \frac{\text{SI}_{\text{LA}}}{\text{M}_{\text{BP}}}
\end{equation}
where $\text{Th}$ is the threshold,  $\text{SI}_{\text{LA}}$ represents the mean signal intensity of the LA wall and  $\text{M}_{\text{BP}}$ is the LA blood-pool mean intensity.

\subsection{Competitions in Scar Segmentation}

This section focuses on two main challenges/competitions in scar segmentation. 

\subsubsection{Left Atrial and Scar Quantification and Segmentation Challenge (LAScarQS 2022)}

\begin{figure}[ht!]
\centering


\begin{subfigure}{\linewidth}
\includegraphics[width=\linewidth]{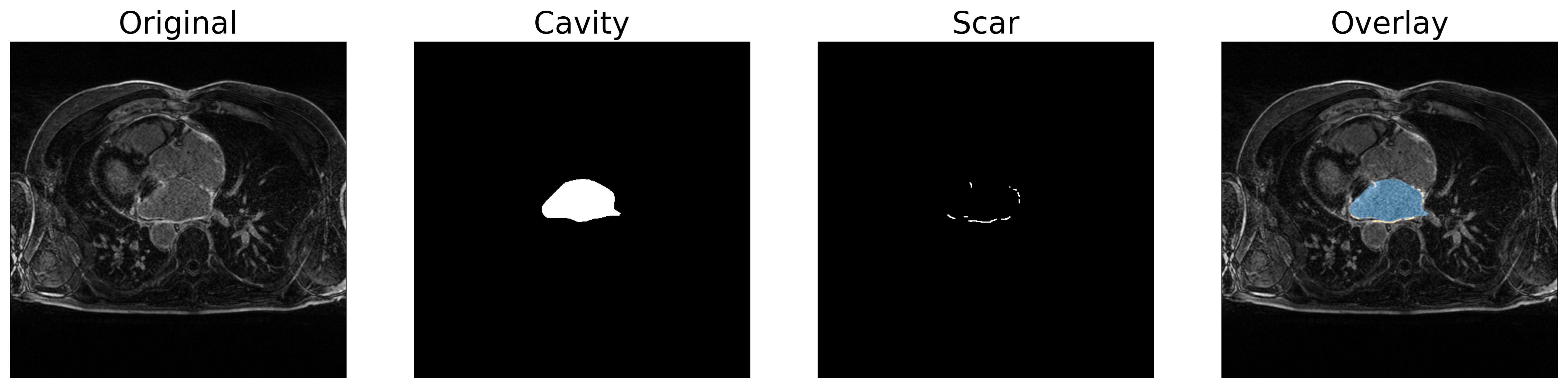}
\end{subfigure}

\begin{subfigure}{\linewidth}
\includegraphics[width=\linewidth]{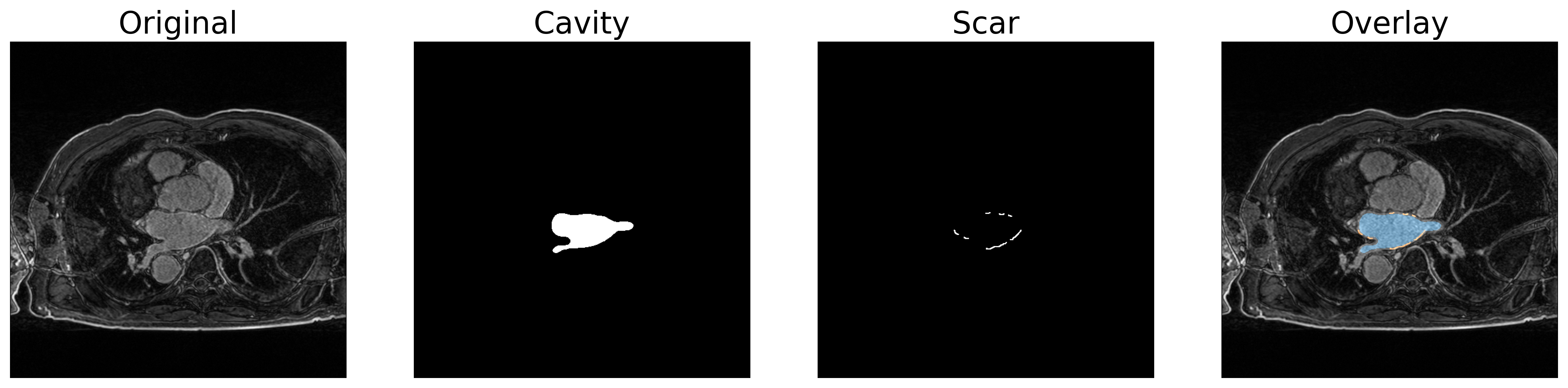}
\end{subfigure}



\caption{Visualization of LAScarQS-2022 images. Each column shows the original image, cavity mask, scar mask and overlay image of the original and masks}
\label{fig:lascarqs}
\end{figure}

Left Atrial and Scar Quantification and Segmentation Challenge \cite{zhuang2023left} (LAScarQS 2022) was held in 2022 in conjunction with leading medical imaging conference MICCAI. This dataset consists of 194 LGE-MRI images taken from three centres around the world.  MRI images acquired from the University of Utah use Siemens Avanto 1.5T or Vario 3T using free-breathing (FB) with navigator-gating. The spatial resolution of one 3D LGE MRI scan was 1.25×1.25×2.5 mm. Beth Israel Deaconess Medical Center extracted the images using  Philips Achieva 1.5T using FB and navigator-gating with fat suppression. The spatial resolution of one 3D LGE MRI scan was 1.4×1.4×1.4 mm. The third and final centre was King’s College London.  Philips Achieva 1.5T using FB and navigator-gating with fat suppression is used to obtain the LGE-MRI images, and the spatial resolution of one 3D LGE MRI scan was 1.3×1.3×4.0 mm. 
The challenge consists of two tasks: Task 1:- segmenting both scars and cavity and Task 2: segmenting cavity only. All submitted frameworks are divided as follows. Sample images are shown in Fig.~\ref{fig:lascarqs}.

\paragraph{One stage networks}

MDBAnet \cite{wu2022multi} is a two-branch network based on a multi-depth strategy. These branches are responsible for segmenting the left atrium (LA) and scars separately. They use the Sobel filter module to transfer information between two branches. This strategy allows to segment of LA scar on the conditions of LA boundary regions. \cite{li2022cross} address the scar segmentation using a different approach compared to other studies: using cross-domain segmentation. Using 3D U-Net as the backbone, they utilize a multi-scale decision-level fusion strategy to achieve competitive performance.  \cite{punithakumar2022automated} propose to use nnU-Net with applying geometric orientations to the inputs to segment LA scars. 

 \underline{Uncertainty based methods}

\cite{zhang2022automatically} and \cite{arega2022using} utilize uncertainty measurement to improve the performance of the models. In \cite{zhang2022automatically}, authors introduce an uncertainty-aware module to lower the threshold of the probability maps in uncertain predictions. They use a variant of nnU-Net to segment scars, and  TopK loss with Dice loss is used to extract information about the boundary. Soft boundary distance maps are utilized to guide the prediction of LA scars. \cite{arega2022using} uses two different losses for LA segmentation and scar segmentation. Scar segmentation utilizes uncertainity information due to its challenging nature. Dice loss with polynomial cross-entropy loss is utilised for LA cavity segmentation.

\paragraph{Two stage networks}

UGformer \cite{liu2022ugformer} is a novel architecture which utilizes graph convolutional networks, transformers and decoders from U-Net to achieve competitive performance in scar segmentation. Utilizing the graph convolutional networks, authors were able to optimize the global space of intermediate feature layers. In this network, a novel transformer block works as the encoder, while convolutional blocks work as the decoder. Graph neural networks with skip connections work as the bridge. LA-HRNet \cite{xie2022hrnet} is another two-stage network which is inspired by the whole brain segmentation network proposed by Li et.al (2021) \cite{li2021whole}. By using feature fusion, authors were able to capture rich features, and they propose feature reuse to improve the quality of the features. Apart from that, their novel auxiliary loss makes the model converge faster. LASSNet \cite{lefebvre2022lassnet}
is another two-stage network which utilizes four neural networks to segment both the LA cavity and scars. In each stage, regions or scars are segmented after selecting the region of interest to segment. \cite{zhang2022two} mainly focuses on the domain shift issue, which can reduce the model's performance. They proposed a histogram matching-based solution to avoid it and trained the model using a two-stage nnU-Net. \cite{khan2022sequential} addresses the LA segmentation and scar quantification jointly using a multi-scale weight-sharing network and boundary-based processing.  TESSLA \cite{ogbomo2022tessla} is a two-stage ensemble deep-learning model which utilizes intensity ratio normalization to segment both LA and scars.

\underline{Utilizing different types of learning methods}

\cite{tu2022self} utilizes a self-supervised learning strategy to learn representations from MRI images to improve the segmentation performance. They have utilized the widely used Mask Autoencoder (MAE) method for this purpose. There are two main stages in self-supervised learning: pertaining and fine-tuning.

\begin{wraptable}{r}{8cm}
\centering
 \vspace{-7mm}
\caption{Average Dice score values from the LAScarQS 2022 challenge. The values above the line represent the validation results, while the values below the line indicate the test set results.}

\begin{tabular}{cccc}
\toprule

\textbf{Paper} & \multicolumn{2}{c}{\textbf{Task 1}} & \textbf{Task 2} \\ \cmidrule{2-4}

 & \textbf{Scar} & \textbf{Cavity} & \textbf{Cavity} \\ \midrule

\cite{punithakumar2022automated}  & \textbf{0.660} & 0.907 & \textbf{0.893} \\
\cite{jiang2022deep}  & 0.641 & 0.902 & 0.890 \\ 
\cite{arega2022using}  & 0.634 & 0.898 & - \\ 
\cite{mazher2022automatic}& 0.602 & 0.875 & - \\ 
\cite{zhang2022automatically}  & 0.598 & 0.880 & 0.890 \\
\cite{lefebvre2022lassnet} & 0.553 & \textbf{0.938 }& 0.846 \\
\cite{li2022cross}  & -     & -     & 0.883 \\ 
\cite{zhang2022two}  & -     & -     & 0.878 \\
\cite{khan2022sequential}& -     & -     &0.886\\
\cite{liu2022ugformer}& -     & -     &0.784\\
\cite{xie2022hrnet}& -     & -     &0.853\\
\cite{wu2022multi}& -     & -     &0.858\\
\cite{zhou2022edge}& -     & -     &0.879\\
\cite{tu2022self}& -     & -     &0.873\\ 
 \hline
    \cite{jiang2022deep}&  \textbf{0.595}&  \textbf{0.947}& \textbf{0.831}\\
     \cite{liu2022ugformer}&  0.582&  -& -\\
     \cite{punithakumar2022automated}&  0.561&  0.908& -\\
     \cite{lefebvre2022lassnet}&  0.549&  0.922& 0.792\\
     \cite{mazher2022automatic}&  0.478&  0.828& -\\
     \cite{li2022cross}&  0.188&  -& -\\
     \cite{khan2022sequential}&  -&  0.940& 0.824\\
     \cite{tu2022self}&  -&  -& 0.803\\

\bottomrule
\end{tabular}
\label{tab: lascar val sorted}
\end{wraptable}
In the pretraining stage, the image is divided into patches. Then those patches are masked randomly, and both masked and unmasked patches are fed to the encoder. Now the decoder is tasked to reconstruct the image. If the decoder is successful, it can be assumed that the model has learnt underlined representations. In the fine-tuning phase, authors utilize a new decoder instead of the pre-trained decoder. \cite{mazher2022automatic} proposed a semi-supervised learning-based segmentation approach using pseudo labelling called as 3DResUNet. They have developed two models: 1) using deep supervision to obtain pseudo labels. 2) Utilizing nnUnet that uses pseudo segmentation labels of the first model as actual labels. \cite{zhou2022edge} propose a novel multi-task learning method which utilises edge information to understand the relationship between scars and LA.  \cite{jiang2022deep} utilizes a curriculum learning strategy with U-Net architecture. It helps to learn difficult scenarios gradually. 


Table \ref{tab: lascar val sorted} shows the performance of each method in the validation and test set for the LAScarQS 2022 challenge. Values are obtained from the challenge website. The highest dice scores are indicated using bold values.

\subsubsection{Left Atrial scar segmentation challenge-2013}

This challenge collects images from the Utah School of Medicine, Beth Israel Deaconess Medical Centre and Imaging Sciences at King's College London. The dataset consists of 60 images (30 pre-ablation and 30 post-ablation images). The resolutions of the images are 1.25×1.25×2.5mm, 1.4×1.4×1.4mm and 1.3×1.3×4.0mm, respectively. Since this is an early challenge, most methods are based on traditional machine learning methods, including thresholding and clustering-based methods. More details of the evaluated methods are available at \cite{karim2013evaluation}. 


\subsection{Unaffiliated Research}

In the early attempts of using deep learning for scar segmentation, \cite{yang2017fully} proposed a fully-automated multi-atlas-based whole heart segmentation method for deriving the geometry of the left atrium and pulmonary veins objectively. This is followed by the implementation of a fully automatic deep-learning method for the segmentation of atrial scarring. The deep learning approach incorporates a feature extraction step, achieved through super-pixel over-segmentation, and a supervised classification step, conducted using a stacked sparse auto-encoder. Later, Yang et al. \cite{yang2020simultaneous} present a novel joint segmentation method utilizing a multiview two-task recursive attention model. It is focused on segmenting the left atrium and scar segmentation jointly and simultaneously. The proposed method primarily comprises a multiview learning network coupled with a dilated attention network.  The multiview learning network captures the correlation between 2D axial slices through a sequential learning subnetwork, while concurrently, two dilated residual subnetworks assimilate complementary information from sagittal and coronal views. These insights are then merged into the axial slice features, resulting in enriched multiview features conducive to precise segmentation. Later, Li et al. \cite{li2020atrial}  introduced an innovative framework utilizing graph cuts, marking significant advancements in the methodology. This framework is notable for two key contributions: Firstly, it presents a novel approach for quantifying LA scarring using the formulation of a method based on a surface mesh. Secondly, the study integrates multi-scale learning with convolutional neural networks (CNNs). However, \cite{li2020joint} argues that using muli-scale CNN and graph-cuts separately, they did not achieve end-to-end training. Hence, \cite{li2020joint} proposes MTL-SESA, an end-to-end multi-task learning network designed for the simultaneous segmentation of the left atrium and quantification of scars. The proposed method stands out by incorporating spatial information directly into the pipeline. A key feature of this approach is the implementation of a spatially encoded loss, which is based on the distance transform map and is achieved without necessitating any modifications to the network architecture itself. Improve upon this work; the same authors proposed the AtrialJSQnet \cite{li2022atrialjsqnet} as an extension to \cite{li2020joint}. In AtrialJSQnet, the spatial relationship between the left atrium and scars is emphasized by adopting the left atrium surface as an attention mask on the predicted scar probability map for shape attention. This approach projects scars onto the left atrium surface, effectively bypassing the complex task of wall segmentation. Additionally, the researchers introduce a novel spatially encoded (SE) loss, which integrates spatial information into the pipeline without altering the network architecture.



\subsection{Associated Challenges in Scar Segmentation}

Scar segmentation has its own challenges compared to other image segmentation (e.g., natural images). In this section, we delve into those in detail. 

\subsubsection{Class Imbalance and loss function}

One of the main challenges in deep learning is class imbalance. In scar segmentation, this issue becomes more severe. As shown in Fig.~\ref{fig:lascarqs}, scars only contain a small fraction compared to the overall image. Hence, the foreground (scar) and background are severely imbalanced. Even in cavity background contains around 99.57\%, and in scar, it is around 99.97\%. Hence, researchers are focused on extracting regions of interest (ROI) to overcome class imbalance \cite{zhou2022edge}. Some authors are focused on cropping the input images to remove unwanted backgrounds \cite{jiang2022deep}. Apart from that some authors argue that traditional loss functions such as Dice Loss or cross-entropy losses are unable to handle class imbalance problem  \cite{khan2022sequential}, and they are much more focused on volume over contours, potentially compromising the learning of accurate edges in favour of achieving correct volume measurements \cite{jamart2020mini}. Hence, authors \cite{khan2022sequential} proposed novel loss functions to alleviate class imbalance. Apart from that \cite{arega2022using} suggests DiceFocal loss can show good performance on an imbalanced dataset compared to the DiceLoss. They proposed to utilize uncertainty to improve the performance in highly imbalanced datasets in conjunction with DiceFocal loss.

\subsubsection{2D vs 3D}

Several factors must be weighed in the debate between using 2D versus 3D models for computational tasks, particularly in terms of performance and computational demands. 2D models are recognized for their efficiency. Hence, we can use fewer GPUs with shorter training time, and it has the ability to process larger batch size, which effectively consumes less memory. Conversely, 3D models excel in capturing the full spatial complexity of data, leveraging the depth and continuity between slices to learn comprehensive spatial characteristics. This depth-aware learning enables 3D models to construct more precise representations of three-dimensional anatomy, often resulting in superior accuracy. Moreover, advancements in GPU technology are gradually diminishing the memory constraints associated with 3D models, making them more feasible for widespread use. As datasets grow in size and quality, the benefits of 3D modelling become increasingly apparent, promising enhanced performance in applications such as clinical imaging deep learning.
Choosing between 2D and 3D models involves navigating a trade-off between the lower memory requirements and quicker training times of 2D models and the superior spatial accuracy offered by 3D models, albeit at the expense of increased computational demands.

\subsubsection{Other challenges}

Apart from the above-mentioned challenges,  MRI images can vary significantly in quality and appearance due to different imaging parameters, patient movement, and the inherent heterogeneity of cardiac tissues. These variations make it difficult to develop a universally applicable segmentation algorithm. Also cardiac scars can vary greatly in size, shape, and location, depending on the extent of the myocardial injury and the stage of healing. This heterogeneity makes it challenging to identify and segment the scar tissue accurately. Apart from that , the contrast resolution of MRI images might not always be optimal for distinguishing scar tissue from healthy myocardial tissue or from other pathologies. Even though LGE-MRI can improve contrast for scar tissue but may still present challenges in cases of diffuse fibrosis or when the enhancement is subtle. Moreover, MRI images often exhibit intensity inhomogeneity which can lead to a non-uniform appearance of tissues across the image. This effect can complicate the segmentation of scar tissues, which might have similar intensities to the surrounding myocardium under inhomogeneity conditions. The partial volume effect, where a voxel contains a mixture of tissues, can lead to inaccuracies in segmentation. This is particularly problematic in the case of thin or irregularly shaped scars.

\section{Discussion and Conclusion}

This comprehensive review has delved into the intricate anatomy of the heart, the pathophysiology of atrial fibrillation, the advent and progression of deep learning technologies, and their application in atrial scar segmentation. A significant focus has been on the exploration of atrial segmentation, where it is observed that despite numerous research efforts, the efficacy of current segmentation methodologies remains suboptimal. This limitation is primarily attributed to the inherent complexities associated with scar characteristics, such as their typically small size and the presence of adjacent enhanced regions, as exemplified in Fig.~\ref{fig:lascarqs}.

Moreover, the quantification of scars in MRI images presents its own set of challenges, predominantly due to the scarcity of available data. This limitation often leads to difficulties in achieving high accuracy and reliability in scar identification and quantification. As a response to these challenges, there has been a growing trend among researchers to pivot towards unsupervised learning methods. These methods, which do not rely on labelled training data, are being increasingly favoured for their potential to navigate the nuances of MRI scar data better. The exploration of unsupervised learning in this context is particularly promising. It offers an avenue to overcome the data limitations by utilizing algorithms capable of learning and identifying patterns within the data without the need for extensive and often unavailable annotated datasets. This approach could lead to more sophisticated and accurate models for atrial scar segmentation, which is crucial for the effective diagnosis and treatment of AFib.

Furthermore, the development of more advanced deep learning models, particularly those tailored to handle the specific challenges of cardiac imaging data, is essential. These models need to be adept at handling the variability and complexity of cardiac structures, as well as the subtle nuances of scar tissues in MRI images. The integration of domain-specific knowledge into these models could further enhance their performance, making them more robust and reliable for clinical applications.

In conclusion, while the field of atrial scar segmentation using deep learning has made considerable strides, there remains significant room for improvement. The transition towards unsupervised learning methods and the development of more advanced, domain-specific models present promising pathways for future research. These advancements could lead to more accurate, reliable, and clinically applicable solutions for the segmentation and quantification of atrial scars, ultimately contributing to better patient outcomes in the management of AFib.

\bibliography{sn-bibliography}

\end{document}